\documentclass[12pt,preprint]{aastex}

\slugcomment{Version of 10 June 2009}

\shorttitle{The Spectral Energy Distributions of late-L and early-T Dwarfs}
\shortauthors{Stephens et al.}

\newcommand\teff{\mbox{$T_\mathrm{eff}$}}
\newcommand\fsed{\mbox{$f_\mathrm{sed}$}}
\newcommand\logg{\mbox{$\log g$}}
\newcommand\microns{$\mu$m}

\begin{document}

\title{The 0.8--14.5 \microns\ Spectra of Mid-L to Mid-T Dwarfs:\\  Diagnostics of
Effective Temperature, Grain Sedimentation, Gas Transport, and Surface Gravity}


\author{D.~C. Stephens\altaffilmark{1}}
\email{denise\_stephens@byu.edu}

\author{S.~K.\ Leggett\altaffilmark{2}}

\author{Michael C.\ Cushing\altaffilmark{3}}

\author{Mark S.\ Marley\altaffilmark{4}}

\author{D.\ Saumon\altaffilmark{5}}

\author{T.~R.\ Geballe\altaffilmark{2}}

\author{David~A.\ Golimowski\altaffilmark{6}}

\author{Xiaohui Fan \altaffilmark{7}}

\and

\author{K.~S.\ Noll\altaffilmark{6}}

\altaffiltext{1}{BYU Department of Physics and Astronomy, N486 ESC, Provo, UT  84602}

\altaffiltext{2}{Gemini Observatory, Northern Operations Center, 670 N.\ A'ohoku Place, Hilo, HI 96720}

\altaffiltext{3}{Institute for Astronomy, University of Hawaii, 2680 Woodlawn Drive, Honolulu, HI 96822}

\altaffiltext{4}{NASA Ames Research Center, MS 245-3, Moffett Field, CA 94035}

\altaffiltext{5}{Los Alamos National Laboratory, PO Box 1663, MS F663, Los Alamos, NM 87545}

\altaffiltext{6}{Space Telescope Science Institute, 3700 San Martin Drive, Baltimore, MD 21218}

\altaffiltext{7}{Steward Observatory, University of Arizona, Tucson, AZ 85721}

\begin{abstract}

We present new 5.2--14.5~\microns\ low-resolution spectra of 14 mid-L to mid-T dwarfs.  We also present new 3.0--4.1~\microns\ spectra for five of these dwarfs.  These data are supplemented by existing red and near-infrared spectra ($\sim$ 0.6--2.5~\microns), as well as red through mid-infrared spectroscopy of seven other L and T dwarfs presented by Cushing et al.\ (2008). We compare these spectra to those generated from the model atmospheres of Saumon \& Marley (2008).  The models reproduce the observed spectra well, except in the case of one very red L3.5 dwarf, 2MASS J22244381$-$0158521. The broad wavelength coverage allows us to constrain almost independently the four parameters used to describe these photospheres in our models: effective temperature (\teff), surface gravity, grain sedimentation efficiency (\fsed) and vertical gas transport efficiency ($K_{zz}$).  The CH$_4$ bands centered at 2.2, 3.3, and 7.65~\microns\ and the CO band at 2.3~\microns\ are sensitive to $K_{zz}$, and indicates that chemical mixing is important in all L and T dwarf atmospheres.
The sample of L3.5 to T5.5 dwarfs spans the range $1800 \gtrsim$ \teff\ $\gtrsim 1000$~K, with an L-T transition (spectral types L7--T4) that lies between 1400~K and 1100~K for dwarfs with typical near-infrared colors; bluer and redder dwarfs can be 100~K warmer or cooler, respectively, when using infrared spectral types.  When using optical spectral types the bluer dwarfs have more typical \teff\ values as they tend to have earlier optical spectral types.  In this model analysis, \fsed\ increases rapidly between types T0 and T4, indicating that  increased sedimentation can explain the rapid disappearance of clouds at this stage of brown dwarf evolution.  There is a suggestion that the transition to dust-free atmospheres happens at lower temperatures for lower gravity dwarfs. 

\end{abstract}


\keywords{
stars: low-mass, brown dwarfs --- stars: individual
(2MASS J00361617$+$1821104,
2MASS J05591914$-$1404488,
2MASS J08251968$+$2115521,
2MASS J09083803$+$5032088,
2MASS J15074769$-$1627386,
2MASS J22244381$-$0158521,
2MASS J22443167$+$2043433,
2MASS J22541892$+$3123498,
DENIS-P J025503.3$-$470049,
SDSS J000013.54$+$255418.6,
SDSS J075840.32$+$324723.3,
SDSS J080531.83$+$481233.1,
SDSS J085758.44$+$570851.4,
SDSS J105213.50$+$442255.6AB,
SDSS J111009.99$+$011613.0,
SDSS J115553.85$+$055957.5, 
SDSS J120747.17$+$024424.8,
SDSS J125453.90$-$012247.5,
SDSS J133148.88$-$011652.5,
SDSS J151643.00$+$305344.3,
SDSS J152039.82$+$354619.8)
}

\section{Introduction}

Brown dwarfs have masses too low to sustain hydrogen fusion in their cores.  Without an internal source of energy, these objects do not maintain a relatively constant luminosity and temperature as main-sequence stars do.  Instead they slowly cool, radiating away the energy produced during their formation and contraction.  As they dim and cool, their spectral energy distributions (SEDs) evolve through progressively later spectral types as the primary sources of opacity change with decreasing effective temperature (\teff). Atomic and molecular species that were present in the atmosphere of the young, warm brown dwarf are gradually replaced by new molecules whose existence is favored at lower temperatures.   Iron and silicate grains begin to condense, providing additional sources of opacity in the form of iron, forsterite (Mg$_2$SiO$_4$), and/or enstatite (MgSiO$_3$) clouds (Lunine et al. 1986, Tsuji et al. 1996, Burrows \& Sharp 1999, Lodders 1999, Marley 2000, Allard et al.\ 2001, Marley et al.\ 2002).  As a brown dwarf ages and \teff\ decreases, the grains condense more deeply in the atmosphere until the cloud layers lie below the photosphere, break apart, or otherwise dissipate (Ackerman \& Marley 2001, Tsuji 2002, Burgasser et al. 2002b, Knapp et al. 2004).  These chemical and dynamical changes in the atmosphere alter the strength and appearance of the spectral features used to classify brown dwarfs.  Consequently, a brown dwarf's spectral type progresses through the M, L, and T classes as it cools.   Informative reviews of ultracool and brown dwarfs have been written by Chabrier \& Baraffe (2000), Burrows et al.\ (2001), Kirkpatrick (2005) and Marley \& Leggett (2008).

The boundaries between the M and L spectral classes and the L and T spectral classes are marked by major changes in atmospheric chemistry and dynamics.  The transition from M to L is best characterized by the disappearance of metal oxide bands and the increasing strength of metal hydride bands as grains form (Kirkpatrick et al.\ 1999).  The transition from L to T is heralded by the appearance of CH$_4$ absorption features centered at 1.6 and 2.2 \microns\ (Geballe et al.\ 2002; Burgasser et al.\ 2002a) and the disappearance of silicate clouds from the photosphere.  The appearance of CH$_4$ and the loss of cloud opacity cause an abrupt change in the near-infrared colors of brown dwarfs.  The increasingly red near-infrared colors ($1.5 \lesssim J$--$K \lesssim 2.0$) of late-L dwarfs become bluer ($J$--$K \sim 0$) as they evolve into early-T dwarfs (Knapp et al.\ 2004).  A brightening in the $J$-band  is also observed (Dahn et al.\ 2002), such that early-T dwarfs are brighter than late-L dwarfs in this bandpass.  This brightening is a real effect and not due to variations in other parameters as verified by several studies of binary systems 
(Gizis et al.\ 2003, Cruz et al.\ 2004, Burgasser et al.\ 2006b, Liu et al.\ 2006, Looper et al.\ 2008).

Whereas the M--L transition is in all likelihood caused by decreasing \teff\
with the associated changes in chemistry and the onset of condensation, 
the widely varying SEDs of the latest L dwarfs suggest that the spectral properties 
of dwarfs near the L--T transition are additionally influenced by the cloud characteristics
and other fundamental properties such as gravity, metallicity, and rotational velocity.  
Kirkpatrick et al.\ (2000) noted a relative scarcity of L--T transition dwarfs and suggested that the L--T transition occurs over a very narrow range of \teff.  Golimowski et al.\
(2004a) and Vrba et al.\ (2004) confirmed this narrow temperature range using values of \teff\ derived from the measured bolometric luminosities of many L and T dwarfs and radii from the evolutionary models of Burrows et al.\ (1997), Baraffe et al.\ (1998), and Chabrier et al.\ (2000).  They found that \teff\ is roughly constant among the late-L to early-T dwarfs (see \S 9).  Thus late-L and early-T dwarfs display spectral diversity over a relatively narrow range of effective temperature.

We have developed atmospheric models and synthetic spectra to better understand the extent to which rapidly changing chemistry and dynamics influence the SEDs of L and T dwarfs, particularly near the L--T transition.  The models span a range of effective temperatures and gravities appropriate for brown dwarfs with solar metallicity.  The models also include two additional parameters, one that accounts for the vertical extent and particle sizes of clouds, and a second to account for the degree to which vertical mixing drives chemical species out of equilibrium.  The first parameter, $f_{\rm sed}$, accounts for variations in cloud properties (Ackerman \& Marley 2001; Marley et al.\ 2002).  Small values of  $f_{\rm sed}$ represent thick clouds with small grain sizes, and large values represent thin clouds with large grain sizes.  We generally find that the colors and spectra of L dwarfs are reproduced with \fsed\ $= 1$--2, while those of later T dwarfs are reproduced with \fsed\ $=4$ or cloud-free ($f_{\rm sed} \rightarrow \infty$) models (Knapp et al.\ 2004; Cushing et al.\ 2008; Saumon \& Marley 2008).

The detailed physics of cloud formation is quite complex and difficult to model.  The complexity cannot be adequately represented by a single parameter such as \fsed, and several groups have adopted various approaches in trying to model this phenomenon.  In the Tsuji Unified Cloud Model (Tsuji 2002; 2005) the grains condense at one temperature and precipitate at a lower temperature.  The difference between these temperatures defines the thickness of the cloud.  The ``Settl'' models of Allard et al.\ (2003; 2007) strike a balance between the growth rates of grains and the mixing time scale by iteratively adjusting the 
condensate fraction or the grain sizes until the former no longer changes.  Woitke \& Helling (2003) and Helling et al.\ (2008a) use seed particles and nucleation to model grain growth and then allow the grains to drift into regions where they may continue growing before settling and evaporating deep in the atmosphere.  A thorough review of the recent approaches to cloud modeling is given by Helling et al.\ (2008b).  Comparing these models is beyond the scope of this paper, but we encourage others to use the data presented here to test their models.

The second additional parameter used in our models is $K_{zz}$ (cm$^{2}$ s$^{-1}$), the eddy-diffusion coefficient in the nominally stable region above the radiative-convective boundary.   This parameter is related to the timescale of vertical transport, and is incorporated to account for how vertical mixing can alter the chemical abundances of C-, N- and O-bearing species.  Stable CO and N$_2$ molecules can be dredged up from the deep atmosphere to the photosphere, causing enhanced abundances of these species and decreased abundances of CH$_4$, H$_2$O, and NH$_3$ (Fegley \& Lodders 1996; Lodders \& Fegley 2002; Saumon et al.\ 2003, 2007; Hubeny \& Burrows 2007).  Generally, larger values of $K_{zz}$ imply greater enhancement of CO and N$_2$ over CH$_4$ and NH$_3$, respectively.  Values of $\log K_{zz} = 2$--6, corresponding to mixing times of $\sim 10$~yr to $\sim 1$~hr, reproduce well the near-infrared spectra of T dwarfs (Leggett et al.\ 2007).

In this paper, we present new 5.2--14.5~\microns\ spectra of 14 late-L and T dwarfs obtained with the {\it Spitzer Space Telescope} (Werner et al.\ 2004).  We use these spectra, augmented with similar spectra of seven mid-L to early-T dwarfs presented by Cushing et al.\ (2008), and our model atmospheres to investigate how differences in the physical characteristics of brown dwarfs will change their SEDs and spectral features at and beyond the L--T transition.  Near-infrared (1.0--2.5~\microns) spectra for these dwarfs already exist, but our mid-infrared spectra provide a better opportunity to untangle the various effects of \teff, gravity, sedimentation efficiency, and vertical mixing over a long flux baseline.  By identifying correlations between these parameters and observed spectral type, we hope to better understand the SEDs of brown dwarfs and to encourage the development of models that more realistically capture the complex atmospheric physics of these dwarfs.  The next sections of the paper describe the selection of targets for observation, the acquisition and reduction of the spectra, the models and fitting procedures, and the 
results of our analysis.

\section{Study Aims and Sample}

Three fundamental questions define this study: Which physical parameters are most closely associated with the observed range of near-infrared colors of late-L dwarfs ($0.9 \lesssim J$--$K \lesssim 2.4$)?  How do these parameters influence the evolution of L dwarfs into early-T dwarfs?  How does vertical mixing influence the mid-infrared spectra of T dwarfs?  This last question is motivated by the mid-infrared colors of T dwarfs measured from the ground (Noll et al. 1997; Golimowski et al. 2004a) and by {\it Spitzer's} Infrared Array Camera (IRAC, Fazio et al.\ 2004; Patten et al.\ 2006; Leggett et al.\ 2007). These mid-infrared colors can only be reproduced by model atmospheres that include vertical mixing.  To address these questions we selected a handful of late-L dwarfs and several T dwarfs for mid-infrared spectroscopic observation.  The selected late-L and early-T dwarfs span the range of near-infrared colors, and the mid-T dwarfs likely undergo strong vertical mixing.

Our observational sample includes six late-L dwarfs with red ($1.9 \lesssim J$--$K \lesssim 2.4$), normal ($J$--$K \approx 1.5$), and relatively blue ($1.1 \lesssim J$--$K \lesssim 1.3$) near-infrared colors; five T0--T2 dwarfs with normal ($0.9 \lesssim J$--$K \lesssim 1.5$) and red ($J$--$K \approx 1.7$) colors; and three T4--T5.5 dwarfs with normal ($-0.1 \lesssim J$--$K \lesssim 0.0$) and relatively red ($J$--$K \approx 0.1$) colors.  These color designations are based on the observed correlations between color and spectral type presented in Figure~3 of Knapp et al.\ (2004), which are defined on the Mauna Kea Observatory (MKO) photometric system (Simons \& Tokunaga 2002; Tokunaga et al.\ 2002). Because the L--T transition is manifested as a reversal in the redward trend of $J$--$K$, it is plausible that the late-L and early-T dwarfs with unusual colors will reveal, or place limitations on, the agent(s) influencing the transition. 

Table~1 provides basic information about the 14 brown dwarfs for which we obtained mid-infrared spectra with {\it Spitzer}.  Henceforth, we refer to these dwarfs by their abbreviated coordinates and refer the reader to Table~1 for the official designation.  We adopt the near-infrared spectral classifications of Geballe et al.\ (2002) and Burgasser et al.\ (2006a) for the L and T dwarfs, respectively.  The far-optical classifications of Kirkpatrick et al.\ (1999) for the L dwarfs are also provided and discussed in \S 9.  Table~1 includes estimates of the distances to the L and T dwarfs based on the relationships between spectral type and absolute $JHK$ magnitudes derived by Liu et al.\ (2006; we use the relationships determined after excluding known binaries, but not possible binaries).  The uncertainties in the distances listed in Table 1 were found using a monte carlo technique and reflect the root-mean-squared errors associated with the relationships of Liu et al.\ (2006), the uncertainty in near-infrared spectral classification, the uncertainties in apparent magnitude, and the range of absolute magnitude and final distance values found from using the different $J$, $H$, and $K$ absolute magnitude relationships.  Table~1 also lists measured proper motions and implied tangential velocities of the dwarfs. Not surprisingly, the velocities suggest that our sample consists of young to old disk dwarfs (e.g. Eggen 1998).

We have augmented our observational sample with seven mid-L to mid-T dwarfs for which near- and mid-infrared spectra have been published previously.  Table~2 lists some basic information about these dwarfs.  Cushing et al.\ (2005) obtained high-resolution ($\lambda/\Delta\lambda \approx 1200$ or 2000), 0.9--2.4~\microns\ spectra of these dwarfs with the SpeX instrument (Rayner et al.\ 2003) at NASA's Infrared Telescope Facility.  Additional medium resolution 2.9--4.1~\microns\ spectra with SpeX ($\lambda/\Delta\lambda \approx 940$) and with the Infrared Camera and Spectrograph (IRCS) on the 8.2~m Subaru Telescope ($\lambda/\Delta\lambda \approx 425$ or 210; Kobayashi et al.\ 2000) were obtained as well.  Cushing et al.\ (2006) then obtained low-resolution ($\lambda/\Delta\lambda \approx 90$), 5.5--38.0~\microns\ spectra of the seven dwarfs using {\it Spitzer's} Infrared Spectrograph (IRS; Houck et al.\ 2004) in the same configuration used for our observational sample (\S3.1).

The near-infrared colors of L and T dwarfs vary significantly among photometric systems (Stephens \& Leggett 2004) and in this paper we refer only to colors based on the MKO system. The $J - K$ colors given in Tables 1 and 2 are taken from the MKO photometry presented by Leggett et al. (2002), Knapp et al. (2004) and Chiu et al. (2006), except for that of DENIS-P J0255$-$47. For this dwarf, MKO $J$--$K$  was determined by transforming
the 2MASS $J$--$K$ colors (Kirkpatrick et al. 2008)  using the relationships determined by Stephens \& Leggett (2004).  The uncertainty in $J - K$ for the sample is 0.04 mag except for  SDSS 0000$+$25 which has an uncertainty of 0.06 mag, SDSS 1110$+$01 which has an uncertainty of 0.07 mag, and DENIS-P J0255$-$47 which has an uncertainty of 0.08 mag.

\section{Data Acquisition and Reduction}

We observed the L and T dwarfs listed in Table~1 with the Infrared Spectrograph (IRS) aboard the {\it Spitzer Space Telescope} as part of Cycle 1 and 2 Guest Observer programs 3431 and 20514.  We used the Short-Low (SL) module to obtain second-order (5.2--7.7~\microns) and first-order (7.4--14.5~\microns) spectra in the standard staring mode, which produces low-resolution spectra with $\lambda/\Delta\lambda \approx 60$--120.  Low-level processing of the spectra was done automatically by version 13.2.0 of the IRS pipeline, yielding Basic Calibrated Data (BCD).  We further cleaned, processed, and extracted the spectra as described in \S3.1.

We combined the calibrated IRS spectra of the objects in Table 1 with published spectra covering optical and near-infrared wavelengths ($0.4 \lesssim \lambda \lesssim 2.5$~\microns).  The near-infrared spectra were obtained with instruments and procedures described by Geballe et al.\ (2002), Knapp et al.\ (2004), and Chiu et al.\ (2006).  Optical spectra are available for nine of the L and T dwarfs, and references for these spectra are given in Table~1.  We also obtained new 3.0--4.1~\microns\ spectra for five targets using the Near InfraRed Imager and Spectrograph (NIRI; Hodapp et al.\ 2003) at Gemini North Observatory.  Details of the Gemini observations are given in \S3.2.  Combined spectra containing all available data for the 14 L and T dwarfs are shown in Figure~1.

\subsection{IRS Spectra from 5.2 to 14.5 \microns}

For all the targets in Table 1, except SDSS~0857$+$57, we used IRS's short-low (SL) module with a ramp duration of 240~s, which is the integration time for one image.  For SDSS~0857$+$57, our brightest target, we used a ramp time of 60~s.   The total exposure time was chosen to provide data that would be sensitive to flux variations at the 20\% 
level due to cloud and chemical phenomena.  We estimated the mid-infrared flux density for each target using near-infrared magnitudes and the theoretical models of Saumon et al. (2003), and used the $Spitzer$ Spec-PET tool to find the number of integration cycles necessary to obtain signal-to-noise (S/N) ratios of 20--25 in each order.  Taking into account the uncertainty in the model flux densities, this gives 3--5 $\sigma$ detections of the expected flux variations.  Table~3 lists the total on-source integration time for each target.  Cycles were repeated after nodding the target along the slit by roughly one- and two-thirds of its length.

We started the IRS reduction using the Basic Calibrated Data (BCD) images.  These are images produced by the IRS pipeline after all instrumental artifacts have been corrected.  We then further reduced the data using scripts and contributed IDL software.  First, a bad-pixel mask was created by flagging pixels that were variable with time, unstable, or not-a-number (NaN) during the exposure.  The bad pixels were repaired by running the images and mask through IRSCLEAN\_MASK version 1.5 (J.\ Ingalls, personal communication), which is available from the Spitzer Science Center.\footnote{http://ssc.spitzer.caltech.edu/archanaly/contributed/irsclean/IRSCLEAN\_MASK.html}  The sky background was removed from each image by subtracting a median sky frame created from all the images of the same spectral order taken at the other nod position.  After subtracting the sky frames, the residual background was determined from the median signal in each image row (excluding those pixels where the spectra or its negative were located) and then subtracted from each row.  After a visual inspection, the images were loaded into the Spectroscopic Modeling Analysis and Reduction Tool (SMART; Higdon et al.\ 2004) for spectrum extraction.

The spectra were extracted and wavelength-calibrated using a fixed-width aperture of three pixels.  For each target, all extracted spectra of the same order and nod position were combined using a mean-clipping algorithm.  IRS observations of the standard star HR~6348 obtained on dates immediately before and after the target observations were similarly extracted and combined.  A spectral template of HR~6348 (G.~Sloan, private communication) was used to determine the relative spectral response of the standard star in each order and nod position.  Each target spectrum was then normalized by the corresponding spectral-response function.  The target spectra in each order were combined, and the two orders were merged to produce a complete spectrum from 5.2 to 14.5 \microns.  The spectrum was then flux calibrated using IRAC channel 4 photometry from Patten et al.\ (2006) and Leggett et al.\ (2007) using the procedure described in Cushing et al.\ (2006; see also Reach et al.\ 2005).

\subsection{Gemini NIRI Spectra from 3.0 to 4.1 \microns}

Using the Gemini North Telescope, we acquired 3.0--4.1~\microns\ spectra of five of the brighter late-L and early-T dwarfs to measure the fundamental CH$_4$ band centered at 3.3~\microns\ and to partially fill the gap between the near-infrared and IRS spectra (Figure~2).  The 3.3 \microns\ band forms higher in the atmosphere than other CH$_4$ features, and so is an interesting probe of both the vertical mixing coefficient and the atmospheric temperature profile above the condensate cloud decks.  The inclusion of these data are important for determining the bolometric luminosity of each L and T dwarf, and for defining the overall shape of the infrared SED.  

The 3.0--4.1~\microns\ spectra were obtained as part of queue programs GN-2005A-Q-23 and GN-2006B-Q-37 at the Gemini North Observatory.  The facility camera and spectrograph, NIRI, acquired the data using the $L$-grism and widest (6-pixel) slit to maximize the flux from each dwarf.  This slit provided sufficient resolution ($\lambda /\Delta\lambda \approx$ 460) for studies of broad molecular bands.  A total of thirty 2~s exposures were obtained at alternating slit positions in an ABBA dither pattern.  This sequence was repeated until S/N~$\approx 20$ was obtained.  Table~3 lists the date of observation and total on-source integration time for each target.  Similar observations of A and early-F stars located near the targets were performed to remove telluric features and determine the instrument response.  Flat fields were obtained using lamps in the on-telescope calibration unit, and bad-pixel masks were obtained using dark frames. 

We used the Gemini NIRI IRAF package for the initial reduction of the target and calibration star spectra. The AB frames were pairwise subtracted and any poor frames were removed.  The subtracted spectra were then coadded and divided by the flat field.  We used the Figaro software package for the final stages of the image reduction.  The coadded spectra were rectified, extracted, and any obvious bad pixels and cosmic rays were removed.  The extracted positive and negative spectra were cross-correlated and aligned to the same wavelength.  They were coadded to create the final spectrum.  We then used the calibration star spectrum to remove the telluric features and perform initial flux calibration.  The final flux calibration was achieved using either ground-based $L^{\prime}$ (Golimowski et al.\ 2004a) or $Spitzer$ IRAC Channel 1 (Patten et al.\ 2006; Leggett et al.\ 2007) photometry. 

\section{Model Atmospheres}

Using model atmospheres developed by members of our team (Ackerman \& Marley 2001; Marley et al.\ 2002; Saumon et al.\ 2003; Saumon \& Marley 2008), we created synthetic spectra for comparison with the observed spectra of the dwarfs listed in Tables~1 and 2.  The synthetic spectra have solar metallicity and span ranges of \teff, \logg, \fsed, and $K_{zz}$ that are appropriate for late-L and T dwarfs:  $900 \leq$ \teff\ $\leq 1800$~K, with $\Delta$\teff\ = 100~K; $g$ = 300, 1000, and 3000 m\,s$^{-2}$ (or \logg\ = 4.477, 5.0, and 5.477 in cgs units); \fsed\ = 1, 2, 3, 4, and $\infty$ (no clouds); and $K_{zz} = 0$ (no vertical mixing), $10^2$, $10^4$, and $10^6\,\rm cm^2\,s^{-1}$.  This parameterization is a simplification of the complex physics of these atmospheres and is a feature of our models; other groups have adopted different modeling strategies (see \S1).

Our models include all the significant sources of gas opacity: the molecular lines of H$_2$O, CH$_4$, CO, NH$_3$, H$_2$S, PH$_3$, TiO, VO, CrH, FeH, CO$_2$, HCN, C$_2$H$_2$, C$_2$H$_4$, and C$_2$H$_6$; the atomic lines of alkali metals Li, Na, K, Rb, and Cs; continuum opacity from H$_2$ collision-induced absorption; Rayleigh scattering by H$_2$, H, and He; bound-free opacity from H$^-$ and H$_2^+$; and free-free opacity from H$^-$, He$^-$, H$_2^-$, and H$_2^+$ (Sharp \& Burrows 2007; Freedman et al.\ 2008).  The CH$_4$ line list is known to be incomplete at wavelengths shorter than 1.6 \microns.   We assume solar system abundances (Lodders 2003) and calculate the equilibrium abundances of the important C, N, and O bearing gases using the formulation of Lodders \& Fegley (2002) as detailed by Freedman et al.\ (2008).  These equilibrium values are then used to determine the abundances of the various gas species in each atmospheric layer during construction of the temperature-pressure profile.  We calculate opacities and generate line profiles in the manner described by Freedman et al.\ (2008).    The line profiles of the red Na and K doublets follow the model of Burrows et al. (2000).

The models assume radiative-convective equilibrium and produce a temperature-pressure profile for each combination of \teff, \logg, and \fsed\ using the radiative-transfer-solution technique of Toon et al.\ (1989).  During this process, condensates of certain refractory elements are included in atmospheric layers where physical conditions favor such condensation.  The vertical distribution and sizes of grains are controlled in the models by \fsed\ (Ackerman \& Marley 2001).   As the temperature-pressure profile converges, the cloud models are continuously updated until a final self-consistent model including \fsed\ is produced. In addition to these ``cloudy'' models, we also create a ``no cloud'' (nc) model
 where grain condensation is maintained, but the opacity of the clouds is ignored.  This represents the limiting case where the cloud particles fall down in the 
atmosphere and evaporate with perfect efficiency as soon as they condense.
Ultimately \fsed\ and the strength of vertical mixing above the radiative-convective boundary depend on the fundamental characteristics of a given dwarf, including \teff, \logg, metallicity, and perhaps rotation rate.  For now we consider the model parameters to be mutually independent, with the goal of understanding how cloud sedimentation and vertical mixing are related to the fundamental properties of ultracool dwarfs.
Finally, we use the evolutionary sequences of Saumon \& Marley (2008) with surface-boundary conditions set by our cloudy (\fsed\ $= 2$) and cloudless model atmospheres to derive the radius, mass, and age of each dwarf for various values of \teff, \logg, \fsed, and $K_{zz}$.

\section{Analysis}

Using the ranges of \teff, \logg, \fsed, and $K_{zz}$ listed in \S4, we created more than 600 synthetic spectra extending in wavelength from 0.8 to 15.0 \microns.  We compared the synthetic spectra with the observed spectra of the 21 dwarfs listed in Tables 1 and 2 using a fitting procedure similar to the one described by Cushing et al.\ (2008).  Because our spectra lack uncertainties associated with the individual flux densities, we did not use the $G$ statistic defined by Cushing et al.   We instead used two other goodness-of-fit statistics, defined as

\begin{equation}
G_1 = \sum_i w_i (f_i-CF_i)^2
\end{equation}

\noindent
and

\begin{equation}
G_2 = \sum_i w_i (f_i-C'F_i)^2 / C'F_i
\end{equation}

\noindent
where $f_{i}$ and $F_{i}$ are the flux densities of the data and model respectively, and $w_{i}$ is the statistical weight of the $i$th pixel.  The constants $C$ and $C'$ give the flux normalization $(R/D)^2$ between the model and the data, where $R$ and $D$ are the brown dwarf's radius and distance, respectively.   

$G_1$ is a basic least-squares statistic, while $G_2$ is similar to the conventional $\chi^2$ statistic.   Unlike the $G$ statistic defined by Cushing et al.\ (2008), $G_1$ and $G_2$ depend on the flux density units ($f_{\nu}$ or $f_{\lambda} \propto \lambda^{2} f_{\nu}$).  For each model, we determine $C$ and $C'$ (and thus $G_1$ and $G_2$) by minimizing $G_1$ and $G_2$ with respect to $C$ and $C'$.  The best fitting models in our parametric grid are those that produce the smallest values of $G_1$ and $G_2$ for a given dwarf. The $G$ values cannot be compared between objects due to object-to-object flux variations and the flux dependence of both $G_1$ and $G_2$, as well
as variations in the number of data points. Furthermore, when fitting a flux distribution over a large wavelength range, the quality of the fit will be impacted by many different issues, making the goodness of fit dependent on both the wavelength coverage of the data and the spectral type of the object.

Cushing et al.\ (2008) noted that the quality of the model fits are often dominated by systematic errors rather than statistical errors.  For example, if $w_i = 1$ for all $i$, then the resulting fits are biased toward those wavelength ranges having the largest number of pixels (1--2.5 and/or 3~\microns, in our case).  We fitted model spectra using both constant weights ($w_i = 1$) and the weights used by Cushing et al. ($w_i \propto$ pixel width in \microns.)  Unlike Cushing et al., we did not neglect the 1.58--1.75~\microns\ wavelength range.  If our models perfectly reproduced the data, than the best-fitting model spectrum would be independent of the chosen units of flux density.  Our models, however, are not perfect, so we determined the best-fitting spectra using the $G_1$ and $G_2$ stats, combined with either uniform or nonuniform weights, and flux units of either $f_{\lambda}$ of $f_{\nu}$.  Fitting $G_1$ and $G_2$ for each combination of weighting and flux density units resulted in a total of eight best-fitting model solutions for each brown dwarf. 

\subsection{Range of Best Fits}

The eight best fits found from minimizing $G_1$ and $G_2$ can tightly constrain the model parameters for some of the observed brown dwarf spectra, but only loosely constrain the parameters for others. Figures~3 and 4 show examples of each case for dwarfs with \teff~$\approx 1400$~K, i.e., near the middle of our sampled range of \teff.  For SDSS~1520$+$35 (infrared spectral type T0), the best-fitting models indicate \teff~$= 1400$~K, \logg~ $= 4.5$--5.0, \fsed~$= 2$, and $K_{zz} = 0$--$10^2$ cm$^2$\,s$^{-1}$.  For 2MASS~0908$+$50 (infrared type L9; optical type L5), the ranges of model parameters are much wider: \teff $= 1300$--1500~K, \logg~$= 4.5$--5.5, \fsed~$= 2$--3, and $K_{zz} = 10^2$--$10^6$ cm$^2$\,s$^{-1}$.  Typically, minimizing $G_1$ and $G_2$ yields $\Delta$\teff~$= 200$~K and $\Delta$\fsed~$= 1$.  Surface gravity is not well constrained and $\Delta \log K_{zz}$ is 2 or 4.  The fits to the spectrum of the very red L3.5 (L4.5 optical) dwarf 2MASS~2224$-$01 is poor and yields a wider range of temperature with $\Delta$\teff~ $ = 400$~K.  

Table~4 lists the range of allowed parameter values found from the eight best model fits to each dwarf, as well as an estimate of the radius found from the scaling parameter C and the distances as given in Tables 1 and 2.  We have restricted the range of allowed values to include only those models that do not violate the radius constraints described below.  For each dwarf, we adopt a single parameter set where each parameter is near the middle of the allowed (equally-likely) range of values, after allowing for the quantization of our model grid.  For those objects with $3~\mu$m spectroscopy we selected the adopted $K_{zz}$ value from the allowed range by eye, as this could be done by isolating that spectral region (which the statistical fit did not do).  The adopted parameter set is used to compare the model SED to the observed SED in  Figures 5--8.   

We can use the scaling parameter $C = (R/D)^2$ to exclude some of the  statistical model fits on physical grounds.  Armed with a photometric or trigonometric distance (Tables 1 and 2), we can derive the radius of the dwarf from $C$. In determining this radius, we assume that all dwarfs are single except for SDSS~1052$+$44AB which is a recently resolved L-T binary (Liu et al.\ in preparation). The impact of binarity on the sample is discussed in \S 6.  A radius can also be derived for each dwarf from structural models (Saumon \& Marley 2008) using the values of \teff, \logg, and \fsed\ given by the atmosphere model. Some of the model fits produce a radius derived from $C$ (hereafter called the ``scaling radius'') that 
is incompatible with the structural radius, even after accounting for uncertainties in distance and a $\sim 5$\% uncertainty 
in the flux calibration.  The radius test can constrain temperature and gravity, especially for those objects with well determined distance.  We excluded fits   that yielded scaling and structural radii that differed by $\ge 3~\sigma$ (a reasonably generous tolerance to avoid excessive restriction of parameter space). The excluded
fits are listed in the notes to Table 4.

Figures~5--8 show the observed and modeled spectra, where the models have the adopted mid-range atmospheric parameters (Table 4).  The models satisfactorily fit all the spectra except for the the very red L dwarf 2MASS~2224$-$01 (L3.5). Excluding 2MASS~2224$-$01, the average deviation between the modeled and observed SEDs shown in Figures 5 to 8 is 10--20\% (where 
the SED in $f_{\nu}$ is sampled every 0.1~$\mu$m in wavelength space); for 2MASS~2224$-$01 the deviation is 80\%. The 
fits to red L dwarfs typically lie at the low end of our \fsed\ domain and it appears that our cloud model is inadequate at this limit.  Exploratory modeling suggests that very red dwarfs may have smaller particle sizes than those calculated by our baseline cloud model, even for \fsed\ = 1.  We will investigate this possibility in a future paper.

To investigate how an individual parameter affects the model SED we additionally fitted the observed spectrum of each dwarf by eye, using the best statistical fits as a starting point.  We found that the model parameters could be finely tuned almost independently by fitting selected regions of the observed spectra and considering the parameters in order of the size of the impact on the SED. Figure~9 illustrates this process for SDSS~1520$+$35 (T0), the dwarf for which the statistical fitting method gave a well-constrained solution (Figure~3).  The red spectrum in Figure~9 is the best fit by eye to SDSS~1520$+$35.  The four blue spectra in Figure~9 illustrate the steps taken to identify this model. We first constrained \teff\ by examining the ratio of the near- and mid-infrared fluxes, as seen in the top two spectra in Figure~9.   Increasing \teff\ by 100~K significantly reduces the mid-infrared flux relative to the near-infrared flux.  Next, we used the slope of the 1.0--2.5~\microns\ flux to determine \fsed\ to $\pm 1$. Increasing \fsed\ produces a shallower 1.0--2.5~\microns\ slope, as indicated by the top and middle spectra in Figure~9.  We then used the strength of the $K$-band flux to constrain \logg\ to $\pm 0.5$.  Decreasing \logg\ produces more $K$-band flux, as seen in the top and second-from-bottom spectra in Figure~9.   Finally, we determined $\log K_{zz}$ to $\pm 2$ from the strength of the CH$_4$ absorption bands centered at 2.2~\microns, 7.65~\microns, and (where $L$-band spectra are available) 3.3 \microns.  In all cases our fit by eye fell within the ranges of parameters given by the statistical fits.   

\subsection{Uncertainty in Fits}

Cushing et al.\ (2008) estimated that the uncertainties in their fitting procedure caused by errors in flux calibration and weighting were $\sigma$ (\teff)~$= 100$~K, $\sigma$ ($\log g$) $= 0.5$, and $\sigma$ (\fsed)~$= 1.0$.  These uncertainties are consistent with the ranges of the parameters obtained in our multiple-fit analysis (Table 4), but the coarseness of our parameter grid may lead to overestimated uncertainties.  The average deviation in the parameters derived for our sample are $\sigma$ (\teff)~$= 70$~K, $\sigma$ ($\log g$) $= 0.2$, $\sigma$ (\fsed)~$= 0.3$ and $\sigma$ ($\log K_{zz}$) $= 1.5$.  We can also estimate the uncertainties incurred from our assumption of solar metallicity.  Our preliminary non-solar metallicity models indicate that $[m/H] = \pm 0.2$ yields uncertainties of $\pm$30~K in \teff, $\pm$0.3~dex in $\log g$, and $\pm$0.3 in \fsed.
  
Additional systematic errors arise from incomplete line lists of molecules that are important sources of near-infrared opacity.   We estimate these uncertainties using the analysis of the early-T dwarf HN~Peg~B by Leggett et al.\ (2008).  HN~Peg~B's \teff\ is similar to those of our sample, and its metallicity, age, and distance are well known so that \teff\ and \logg\ can be accurately determined.  Its near-infrared flux distribution is almost identical to that of SDSS~1254$-$01 (Figure~3 of Leggett et al.\ 2008), and the near- to mid-infrared colors of these dwarfs differ 
by $\sim 25$\%.
The differences between the derived model parameters for these dwarfs may therefore reflect systematic biases in the less constrained fits presented here, due to the known inadequacies in  our model spectra.  These differences are $\Delta$\teff~$= 90$~K, $\Delta \log g = 0.2$, $\Delta$\fsed~$= 0.0$, and $\Delta \log K_{zz} = 1.0$.  

Combining the average deviation in the parameters derived for our sample, the uncertainty involved with using only solar metallicity models, and the comparison with HN~Peg~B, the total uncertainty in our derived parameters is estimated to be $\sigma$ (\teff)~$\approx 120$~K, $\sigma$ (\logg)~$\approx 0.4$, $\sigma$ (\fsed)~$\approx 0.4$, and $\sigma$ ($\log K_{zz}$) $\approx 1.8$.  We increase the uncertainty in \fsed\  to 0.5, as a 
minimum systematic error is likely to be half the model grid step size (the adopted uncertainties in the other parameters are greater than half the step size.)
These total uncertainties are representative of the sample as a whole except for 2MASS~2224$-$01. Given the poor fit to this dwarf and the associated uncertainty in the flux scaling, the radius exclusion may not be justified and so we estimate   $\sigma$ (\teff)~$\approx 200$~K for 2MASS~2224$-$01.

\section{Binary Systems}

Of the dwarfs in Table 1, the late-L dwarf SDSS~0805$+$48 is a suspected binary because weak CH$_4$ absorption is present at 1.6~\microns\ but not at 2.2~\microns\ (Burgasser 2007a), and
the T0.5 dwarf SDSS~1052$+$44 has recently been resolved as a binary brown dwarf (Liu et al.\ in preparation).    For these objects we investigated possible multiplicity by fitting each SED with pairs of model spectra.  We obtained a ``best'' composite spectrum by trial 
and error using the known effects of the model parameters illustrated in Figure~9.  We determined the fluxes of the two model spectra using the photometric distance, and the radii inferred from evolutionary models based on the values of \teff\ and \logg\ from the component spectra.  The values of \logg\ are constrained by the assumption of coevality; for each system, \teff\ and \logg\ must be consistent with a single age indicated by the isochrones shown in Figures 4 and 5 of Saumon \& Marley (2008).  

The dwarfs in Table 2 have accurate distances determined by parallax measurements and we can investigate the possibility of multiplicity using the radii implied by our model fits. This test is different from the evolutionary radius test applied in Section 5 and cannot be applied to objects in Table 1 with photometric distances: a test for over-luminosity is invalid if the distance is determined by the luminosity. Of the dwarfs in Table 2, 2MASS~0559$-$14 (T4.5) and SDSS~1254$-$01 (T2) are suspected binary systems because they are anomalously bright in color--magnitude and color--spectral type diagrams (Dahn et al.\ 2002, Liu et al.\ 2006, Cushing et al.\ 2008). We find that 2MASS~0559$-$14 and another dwarf in Table 2, 2MASS~0036$+$18 (L4),
have acceptable composite fits that would require that they are approximately equal-luminosity binaries, however, these fits are not superior to the single solution. The other objects  --- 2MASS~2224$-$01, 2MASS~1507$-$16, 2MASS~0825$+$21, DENIS~0255$-$47, and SDSS~1254$-$01 --- cannot be similar-luminosity binary systems according to our fits. They are either single dwarfs or very unequal binary systems.

Figure~10 shows the modeled and observed SEDs of the four systems that can be fit as binaries, and 
Table~5 summarizes their binary properties.
We now discuss these four and the suspect binary SDSS~1254$-$01 (T2) in order of spectral type.  

\begin{itemize}

\item
2MASS~0036$+$18 (L4) has relatively blue near-infrared colors.  Our model fits indicate that it 
could be a similar-luminosity binary system with
parameters (\teff, \logg, \fsed, $K_{zz}$) = (1600--1700~K, 5.5, 2--3, 0--$10^2$ cm$^2\,s^{-1}$) and a radius of $0.08~R_{\odot}$.  However the 
dwarf is unresolved in {\it Hubble Space Telescope} images (Reid et al.\ 2006) and our single- and composite-model spectra are sufficiently similar to preclude a preferred solution.

\item
SDSS~0805$+$48 (L9.5) is a suspected unresolved binary (Burgasser 2007a) with unusually blue near-infrared colors (\S 9).  We find that its red through mid-infrared spectrum can be matched with a composite 1600~K + 1000~K model.  These temperatures suggest component infrared spectral types of L4.5 and T6, or optical types of L5 and T6 (\S 9), consistent with the optical L4.5 and T5 spectral types inferred by Burgasser (2007a). This object is matched equally well as both a single or a binary object, and a parallax measurement would be useful for further analysis.

\item
SDSS 1052$+$44AB (T0.5) has been resolved as a close L+T binary dwarf by Liu et al.\ (in preparation) using adaptive optics imaging.  We are able to closely match the observed spectrum with a composite 1400~K + 1100~K model.  The spectroscopic distance inferred from the composite model is consistent with the photometric distance estimate in Table 1.

\item
SDSS~1254$-$01 (T2) appears overluminous in color-magnitude diagrams, but it is presently unresolved (Burgasser et al. 2006b).  We are unable to 
obtain good fits using higher-gravity (smaller radius) models that are required for an equal-mass binary system.  Its 
derived radius is consistent with the model parameters of a single dwarf, so its overluminosity probably reflects relative youth (\S 7).  However,
the presence of a companion with significantly lower luminosity cannot be excluded.

\item
2MASS~0559$-$14 (T4.5) also appears overluminous in color-magnitude diagrams, but it is presently unresolved (Burgasser et al. 2003).  Its observed
spectrum can be successfully fitted as a binary system of equal luminosity, in which each component has model parameters (\teff, \logg, \fsed, 
$K_{zz}$) = (1200~K, 5.5, 4, $10^2$--$10^4$ cm$^2\,s^{-1}$) and radii of $0.08~R_{\odot}$.  However, the single- and composite-model spectra are 
not sufficiently satisfactory (Figure~10) to determine whether 2MASS~0559$-$14 is a young, single dwarf or an older, binary system (\S 7).

\end{itemize}

\section{Masses and Ages of the Dwarfs}

Using our evolutionary models and adopted values of \teff\ and \logg, we can estimate the masses and ages of the dwarfs in our sample.  Table~5 lists the implied masses and ages of the four possible binary systems identified in \S6.  Table~6 lists these values for 19 other dwarfs in our sample assuming that they are single.  Table~6 excludes 2MASS~2224$-$01, for which a good model fit was not obtained, and SDSS~1052+44AB, which direct imaging has revealed to be a binary system.  The systematic uncertainties in \teff\ and \logg\ (\S5) were considered when deriving the evolutionary parameters.

Tables 5 and 6 also list the tangential velocities of our sample dwarfs.  The range of velocities is relatively small.  The dwarfs having the largest and smallest velocities are statistically likely to be among the older and younger dwarfs in our sample, respectively, but their ages
are not well constrained  by our spectral analysis.   Six of the 19 dwarfs listed in Table 6 (i.e., 30\% of the sample) appear to have ages 
$\lesssim 1.0$~Gyr.   This percentage is consistent with the results of Eggen (1996), who used Ca~II fluxes to determine that $\sim 30$\% of 
lower-mass-main-sequence stars within 14~pc of the Sun have ages younger than or similar to the Hyades ($\sim 0.6$~Gyr).

\section{Vertical Mixing in L and T Dwarf Atmospheres}

Vertical mixing is expected to have a significant impact on the carbon chemistry of a brown dwarf's atmosphere and consequently its SED.  Vertical mixing transports CH$_4$ molecules from the upper atmosphere to deeper, hotter layers where chemical reactions favor the break down of CH$_4$ to produce CO.  Likewise, vertical mixing pulls hotter gas, rich in CO, into the upper atmosphere where the CO molecules are stable for a long period of time due to the large binding energy.  This results in an non-equilibrium abundance ratio of CO to CH$_4$ in the upper atmosphere (Saumon et al. 2003).  The enhanced abundance of CO yields less flux at 5~\microns, as revealed by ground- and space-based 5~\microns\ observations (Noll et al.\ 1997; Leggett et al.\ 2002; Golimowski et al.\ 2004a; Patten et al.\ 2006; Leggett et al.\ 2007; Geballe et al.\ 2009).  The lower abundance of CH$_4$ produces weaker absorption bands at 2.2, 3.3, and 7.65~\microns.

In their analysis of the 1.0--14.5~\microns\ spectra of L and T dwarfs, Cushing et al.\ (2008) used model atmospheres that did not include vertical mixing, i.e., the photospheres were in chemical equilibrium with $K_{zz}=0$.  However, vertical mixing should have a strong impact on the SEDs of these dwarfs, particularly at 2.0--5.0~\microns\ where CH$_4$ and CO bands are located.  Figure~11 shows the best-fitting synthetic spectra with and without vertical mixing for a mid-L dwarf and an early-T dwarf that were part of the Cushing et al. study.  For both objects, the model with vertical mixing reproduces the 2.0--4.1~\microns\ region of the observed spectrum better than the model without mixing.  Models that include vertical mixing also better fit the 7.65~\microns\ CH$_4$ band, although the effect is less dramatic than for the band at 3.3~\microns\ (Figures~9 and 11).

We rederived the atmospheric parameters of the seven dwarfs studied by Cushing et al.\ (2008) in Table 2 using models that include vertical mixing.  Our values of \teff\ match those of Cushing et al.\ within the combined uncertainties, except in the cases of the very red L dwarf 2MASS~2224$-$01, for which the discrepancy is 300~K, and two other dwarfs for which our values of \teff\ are $\sim 150$~K cooler.   The large discrepancy for 2MASS~2224$-$01 is not surprising given the poor quality of the model fits to this unusual object (\S5).  The smaller discrepancies may reflect 
the need to decrease the CH$_4$ abundance (by increasing \teff) in models that lack vertical mixing.  Increasing \teff\ may require changes to \fsed\ and \logg\ to better match the modeled and observed spectra.  Our values of \fsed\ for all seven dwarfs match those of Cushing et al.\ within the combined uncertainties.  Our values of \logg\ for 2MASS~1507$-$16 and 2MASS~0559$-$14 differ from those of Cushing et al.\ by more than 0.5 dex.  In these cases, the values of \logg\ derived by Cushing et al.\ imply radii that are inconsistent with structural arguments (\S5).

Because our observations and analysis strongly suggest that vertical mixing occurs in all L and T dwarf atmospheres, we believe that our model parameters are superior to those of Cushing et al.\ (2008).   For our sample of L3.5 to T5.5 dwarfs ($1800 \gtrsim$ \teff\ $\gtrsim 1000$~K) 
vertical mixing is significant ($K_{zz} \approx 10^4$ cm$^{2}$ s$^{-1}$) with apparently no dependence on spectral type.   In the future full dynamical models must be generated to understand the physical processes and the interplay between the vertical transport of the gas and the sedimentation of the grains.

\section{The L and T Dwarf Temperature Scale}

Golimowski et al.\ (2004a) and Vrba et al.\ (2004) used trigonometric parallaxes to determine the bolometric luminosities and hence \teff\ for a sample of L and T dwarfs.  Some of these dwarfs have since been found to be binary, and other dwarfs with well determined distances are now known.  We have recalculated \teff\ for the Golimowski et al.\ sample after revising the luminosities of the binaries Kelu-1 (Liu \& Leggett 2005), LHS~102BC (Golimowski et al.\ 2004b), SDSS~J042348.57$-$041403.5 (Burgasser et al.\ 2005), and SDSS~J102109.69$-$030420.1 (Burgasser et al.\ 2006b).  We have also added the recently discovered companion T dwarfs HD 3651B (Mugrauer et al.\ 2006; Burgasser \ 2007b, Luhman et al.\ 2007; Liu et al.\ 2007) and HN Peg B (Luhman et al.\ 2007; Leggett et al.\ 2008) to the sample.

Figure 12 shows our revised version of Figure~6 of Golimowski et al.\ (2004a).   For those dwarfs lacking well constrained ages, the data points reflect \teff\ at a nominal age of 3~Gyr and the error bars reflect the range of \teff\ for ages 0.1--10~Gyr.  (For a given luminosity, younger dwarfs are cooler because their radii are larger.)  The dotted curve corresponds to the original relationship between \teff\ and spectral type found by Golimowski et al. (2004a).  The solid and dashed curves are our revised fifth order polynomial fits to the data using the infrared spectral types over different ranges. For M6 to T8 types the dashed curve describes:
\begin{equation}
T_{\rm eff}=
2265.9 + 347.82 t_{IR} - 60.558 t_{IR}^2 + 3.151 t_{IR}^3 - 0.060481 t_{IR}^4 + 0.00024506 t_{IR}^5
\end{equation}
where $t_{IR}=6,...,9,10,11,...,19,20,21,...$ for types M6,..., M9, L0, L1,...,L9, T0, T1,..., respectively.
We have also calculated a fit to the sample using the optical types for the L dwarfs and infrared types for the T dwarfs, the fit then is:
\begin{equation}
T_{\rm eff}= 4400.9 - 467.26 t_{OPT/IR} + 54.67 t_{OPT/IR}^2 - 4.4727 t_{OPT/IR}^3 + 0.17667 t_{OPT/IR}^4 - 0.0025492 t_{OPT/IR}^5
\end{equation}
For L3 to T8 types, using the same numerical type notation and infrared L spectral types, the solid curve describes:
\begin{equation}
T_{\rm eff}=
44898 - 10560 t_{IR} + 1031.9 t_{IR}^2 - 50.472 t_{IR}^3  + 1.2315 t_{IR}^4 - 0.011997 t_{IR}^5
\end{equation}
If a numerical type translation of $t_{IR}=0,1,2,...,9,10,11,...$ for types L0, L1, L2,...,L9, T0, T1,...
is preferred, then the L3 to T8 relationship becomes:
\begin{equation}
T_{\rm eff}= 3132.7 - 737.32 t_{IR} + 136.66 t_{IR}^2  -  13.21 t_{IR}^3 +  0.63162 t_{IR}^4 - 0.011997 t_{IR}^5
\end{equation}
Note that the last two relationships are only valid for the L3 -- T8 spectral range. 
The scatter in both the fits is $\approx 100$~K.  Figure 12 shows that these fits generally imply cooler temperatures than the Golimowski et al. relationship, especially for early-type T dwarfs where the difference
is 100 -- 200~K.

Figure 13 shows the values of \teff\ for our sample as a function of spectral type, derived by our SED model fitting. The resolved binary SDSS~1052$+$44AB is excluded as the individual spectral types are not available.  Other possible binaries -- 2MASS~0036$+$18 (L4), SDSS~0805$+$48 (L9.5), and 2MASS~0559$-$14 (T4.5) -- are shown as ringed symbols.  Our model derived values of \teff\ are consistent with the luminosity-based temperatures of Golimowski et al.\ (2004a) for the six dwarfs common to both studies, but our values are systematically lower by 100--200~K than those of Golimowski et al.   (The discrepancy for the anomalously red and hard-to-fit  L dwarf 2MASS~2224$-$01 is 300~K.)  This trend suggests that the six dwarfs are younger than the 3~Gyr nominally adopted by Golimowski et al., which is plausible for a local sample that is biased toward brighter (i.e. younger) brown dwarfs.  The temperatures computed for our sample generally follow the \teff\ versus infrared spectral type relationship described by equations 5 or 6, which is shown as a curve in the top panel of Figure 13.

The lower panel of Figure 13 uses optical spectral types for the L dwarfs in our sample. Three of the eleven L dwarfs in our sample have optical and infrared spectral types that differ by more than one subclass. One object does not have an optical type,  SDSS~1155$+05$, and another, SDSS~1207$+$02,  is classified as an infrared T0 and an optical L8; however, the optical scheme terminates at L8 and so the difference for this dwarf is not significant. Of the three significantly different L dwarfs, two were recognized by Knapp et al. (2004) to be unusually blue in the near-infrared: SDSS~0805$+$48, which is classified as an L9.5 in the infrared and an L4 in the optical, and SDSS~1331$-$01, which is L8 in the infrared and L6 in the optical. The former object may be binary as discussed in \S 6. The third dwarf that differs significantly in type is 2MASS~0908$+$50, which is classified as an L9 in the infrared and an L5 in the optical; this dwarf does not have unusual infrared colors but is one of the dwarfs in our sample for which the model statistical fits gave more varied results (Figure 4).  The curved line in the lower panel represents equation 4, the temperature/type relationship calculated using optical spectral types for the L dwarfs.

As described in Stephens (2003) and Knapp et al. (2004, and references therein), the flux from an L dwarf can emerge from either above or below the condensate cloud decks, depending on the wavelength used.  The far-red and 2 \microns\ flux emerges from above the cloud decks and therefore may be better indicators of temperature than the 1.2 and 1.6  \microns\ flux, which emerges from below the cloud decks and which may therefore be a better indicator of cloud opacity. The infrared spectral type relies heavily on the 1.2 -- 1.6 \microns\ region, and less on the  2 \microns\ region. Knapp et al. show that the L dwarfs with unusual near-infrared colors have spectral indices that are more scattered than the other L dwarfs; for the blue L dwarfs the optical and 2 \microns\ indices give earlier types than the indices derived from the 1.2 -- 1.6 \microns\ region (the 2 \microns\ classification for SDSS~0805$+$48 is L6.5, for SDSS~1331$-$01 it is L6, and for 2MASS~0908$+$50 it is L7.5). The lower panel of Figure 13 uses optical L dwarf types and does show less scatter than the upper plot, primarily due to the move to earlier type for SDSS~0805$+$48 and SDSS~1331$-$01.  Using optical types, the blue L dwarfs have temperatures closer to the spectral type average, although the excessively red dwarfs still appear to be $\sim~100$~K cooler than average. This remaining dependency in color implies that the cloud decks are impacting the emergent optical flux to some degree.

Figure 14 shows \teff\ as functions of the colors of the dwarfs in the $J$, $H$, and IRAC bands.  It supports our observed correlation between \teff\ and the ratio of the near- and mid-infrared fluxes (\S 5.1), as well as the suggestion by Warren et al.\ (2007) that the $H$--[4.49] color is a good indicator of \teff.  While the distribution of colors involving $J$ broadens substantially below 1500~K (presumably due to the effects of condensates), those involving $H$ are more tightly confined, except for $H$--[3.55].  The $H$--[4.49] and, to a lesser degree, the $H$--[7.87] colors appear to be good indicators of \teff.  The [4.49] passband will remain viable as {\it Spitzer's} cryogen begins to wane, so IRAC will continue to make significant contributions to studies of the coolest dwarfs.

\section{Grain Sedimentation and the L -- T Transition}

Figure 15 shows how \fsed\ varies as a function of spectral type, where we again have excluded the resolved binary SDSS~1052$+$44AB. The top panel uses infrared types for the L dwarfs while the lower panel uses optical types. In this case we expect the infrared types to follow trends with condensates more closely than the optical types (see \S 9); however, both panels give similar results.  Typical mid- to late-type L dwarfs have \fsed\ $\sim 2$ to 3. L dwarfs with redder infrared colors have \fsed\ $\sim 1$ to 2. Bluer than average L dwarfs have \fsed\ $\sim 3$ in our sample, and Burgasser et al.\ (2008) recently completed a study of the very blue ($J$--$K = 1.1$) L4.5 dwarf 2MASS~J11263991$-$5003550 using the same models, finding \fsed\ $=4$. Using our treatment of the condensate cloud decks (see \S 1) for a fixed spectral type, the bluer dwarfs are best fit by models with less cloudy atmospheres (higher \fsed) and the redder dwarfs by cloudier models (lower \fsed), i.e.  differences in cloud opacity account for the $J$--$K$ color variations of L dwarfs.  In general \fsed\ increases from $\sim 2$ to $\sim 4$ between types T0 and T2, and the photospheres are essentially cloud-free beyond type T4.   Cushing et al.\ (2008) found a similar trend.

If the L to T transition is described in terms of the spectral types L7 to T4, then the transition lies between 1400~K and 1100~K for dwarfs with typical near-infrared colors.  Figure 16 shows
\fsed\ as a function of \teff; here the transition can be thought of as a decrease in cloudiness (increase in \fsed). (The two components of SDSS~1052$+$44AB are included in this plot using the parameters from Table 5, other parameters are taken from Table 4.) It is interesting to note that 
all of the objects in our sample that are still cloudy at low temperatures (i.e. have \fsed\ $\lesssim$3 at \teff\ $<$ 1300~K) are low gravity --- SDSS~0857+57 (L8), SDSS~1516+30 (T0.5), and SDSS~0758+32 (T2) --- except for 2MASS~2244+20 and 2MASS~0825+21 which are unusually reddened, and therefore cloudy, L6--L7.5 dwarfs (Tables 1, 2 and 4). There is thus a suggestion in our small sample that the clearing of the cloud decks occurs at \teff\ $\sim$1300~K for \logg\ $=$ 5.0 and $\sim$1100~K for \logg\ $=$ 4.5; a gravity dependence in the L to T transition has also been suggested by Metchev \& Hillenbrand (2006), Saumon \& Marley (2008) and Dupuy, Liu \& Ireland (2009).

In this model analysis, sedimentation efficiency increases rapidly between T0 and T4, meaning that 
increased sedimentation can explain the rapid disappearance of clouds at this stage of brown dwarf evolution.  Color magnitude diagrams support this interpretation of our models; the brightening in the $J$-band and the change in near-infrared color across the L to T transition can be reproduced by changing \fsed\ while keeping \teff\ constant at $\approx$1300~K, as illustrated in Figures 8 and 9 of Knapp et al. (2004).  Saumon \& Marley (2008) present a detailed study of color magnitude diagrams using evolutionary models where \fsed\ increases as a function of decreasing \teff\ between 1400 and 1200~K. These models successfully reproduce observations. However, as Saumon \& Marley  write, while the transition can be modeled successfully by varying \fsed\, it is merely a parametric description using a specific and simple cloud model. Full hydrodynamic models are required to properly understand the mechanism that causes variations in \fsed\ among otherwise similar dwarfs.  We note that rotational velocities have been reported by Zapatero Osorio et al. (2006) and Reiners \& Basri (2008) for seven L dwarfs in our sample (Table 2), but no correlation between \fsed\ and rotation is apparent in this small sample. Nor is a correlation between \fsed\ and $K_{zz}$ seen in our sample. There does seem to be a hint of a gravity dependence in the cloud clearing, as described above.

\section{Future Work}

In the future, the redder and bluer dwarfs in our sample should be investigated for non-solar metallicity, although a recent investigation by Burgasser et al.\ (2008) suggests that subsolar metallicity alone cannot explain the spectral peculiarities of all unusually blue L dwarfs.
A preliminary investigation of the effect of non-solar metallicity on our 1300 -- 1700~K models indicates that low-metallicity dwarfs have significantly bluer 1 -- 4 \microns\ colors, and metal-rich dwarfs have redder 1 -- 4 \microns\ colors, than solar-metallicity dwarfs.  The 5 -- 15 \microns\ spectrum does not change as much as the shorter wavelength spectrum.  These effects mimic those of changing \fsed, so an even larger grid of metallicity-dependent models must be generated to identify the spectral regions that allow the effects of \fsed\ and metallicity to be untangled. 

Cushing et al.\ (2006) suggested that models that also contain a population of smaller grains above the main cloud deck may do a better job of reproducing the 9 -- 11~\microns\ spectra of the redder L dwarfs.  The poor model fit to the  very red L3.5 dwarf in our sample may also be an indicator of small grains in the clouds (\S 5).   These lines of evidence suggest that a second-generation cloud model is needed to fully capture the diversity of clouds in brown-dwarf atmospheres.

\section{Conclusions}

We have used the models of Saumon \& Marley (2008) to generate and fit synthetic spectra to the red through mid-infrared spectra of 21 L and T dwarfs.  These dwarfs have spectral types L3.5 to T5.5 and some have unusual near-infrared colors.  The models generally reproduce the observed spectra well, which is remarkable given the complex chemistry and gas dynamics of the dwarfs' atmospheres.  Four model parameters -- the usual effective temperature (\teff) and  surface gravity (\logg) along with our grain sedimentation efficiency (\fsed) and vertical gas transport diffusion coefficient ($K_{zz}$) -- were varied in the process of fitting each spectrum.  The wide wavelength coverage of our spectra allows us to constrain these parameters almost independently.  For mid-L to mid-T dwarfs, the parameter that most influences the SEDs is \teff, which determines the ratio of the near- and mid-infrared fluxes.   The slope of the near-infrared flux distribution is highly influenced by \fsed, while the strengths of the  CH$_4$ bands centered at 2.2, 3.3, and 7.65 \microns\ are sensitive to $K_{zz}$.  Finally, for an assumed metallicity, \logg\ can be constrained by the 2~\microns\ flux.

The fit to the spectrum of the very red L3.5 dwarf  2MASS~2224$-$01 is poor.   Future models will explore the impact of smaller grains on the synthetic spectra.  These models will also include non-solar metallicities (as a fifth parameter) and will be compared with the SEDs of dwarfs with unusually blue or red near-infrared colors, as well as our currently best fitting solar-metallicity models.

The L -- T transition is best described in the context of our one-dimensional homogeneous surface models as a change in \fsed; i.e. the bluer colors of the T dwarfs require a rapid increase in condensate sedimentation.  The physical explanation for this rapid increase requires full hydrodynamic models that include grain sedimentation and vertical gas transport in a consistent and detailed fashion. 
Our spectra indicate that vertical gas transport plays a large role in the atmospheres of L and T dwarfs, and occurs in every dwarf in our sample.

We have revised the L and T dwarf \teff\ scale of Golimowski et al.\ (2004a) to account for dwarfs subsequently found to be binary and recently discovered T dwarfs whose luminosities are well determined. 
We confirm the suggestion of Warren et al.\ (2007) that $H$--[4.49] is a good indicator of \teff.
The values of \teff\ determined for our sample agree with the revised \teff\ versus spectral type relationship for field dwarfs, and  the L7 to T4 types  span a narrow range of \teff\  of 1100 -- 1400~K. There is some indication that the cloud decks of the lower gravity dwarfs clear at a lower temperature than the higher gravity dwarfs.  

The quantity and quality of our spectra, as well as the reliability of our complex model atmospheres, represent huge advances in substellar astronomy over the last few years.   The observational advances are largely due to the success of the {\it Spitzer Space Telescope}.  The models should continue to improve as the dependencies on metallicity and grain sizes are investigated.  We are well on our way to understanding the link between the stars and planets.

\acknowledgments

This work is based on observations made with the \textit{Spitzer Space Telescope} and the Gemini North Observatory through programs GN-2005A-Q-23 and GN-2006B-Q-37.  {\it Spitzer} is operated by the Jet Propulsion Laboratory, California Institute of Technology, under a contract with NASA.  Gemini Observatory is operated by the Association of Universities for Research in Astronomy, Inc., under a cooperative agreement with the NSF on behalf of the
Gemini partnership: the National Science Foundation (United
States), the Science and Technology Facilities Council (United Kingdom), the
National Research Council (Canada), CONICYT (Chile), the Australian Research Council
(Australia), Minist\'{e}rio da Ci\^{e}ncia e Technologia (Brazil) and Minist\'{e}rio da Ci\^{e}ncia 
Tecnologia e Innovacion Productiva (Argentina).  
This work has been supported in part by funds from the {\it Spitzer} Cycle~1 and 2 Guest Observer Programs 3431 and 20514 granted through the California Institute of Technology.  S.~Leggett's and T.~Geballe's research is supported by Gemini Observatory.  D.~Saumon's contribution was supported by a $Spitzer$ Cycle 3 Theory grant.  D.~Stephens acquisition and reduction of Gemini data was funded by NASA Grant NAG5-13127.  This publication makes use of data from the Two Micron All Sky Survey, which is a joint project of the University of Massachusetts and the Infrared Processing and Analysis Center, and funded by the National Aeronautics and Space Administration and the National Science Foundation, the SIMBAD database, operated at CDS, Strasbourg, France, and NASA's Astrophysics Data System Bibliographic Services.


{\it Facilities:} \facility{Gemini:Gillett (NIRI), Spitzer}

\clearpage

\begin{deluxetable}{lccccccccc}
\rotate
\tabletypesize{\footnotesize}
\tablecaption{L and T dwarfs newly observed with {\it Spitzer} IRS}
\tablewidth{600pt}
\tablehead{
\colhead{} & \colhead{Epoch\tablenotemark{a}} & \colhead{Sp.~Type\tablenotemark{b}} & \colhead{$J$--$K$\tablenotemark{c}} & \colhead{Relative} &
\colhead{$D$\tablenotemark{e}} & \colhead{$\mu$\tablenotemark{f}} &
\colhead{PA\tablenotemark{f}} & \colhead{$v_{\rm tan}$} & \colhead{} \\
\colhead{Name\tablenotemark{a}}     & \colhead{YYYYMMDD} & \colhead{(IR/Opt)} & \colhead{(MKO)}  
& \colhead{color\tablenotemark{d}} & \colhead{(pc)} & \colhead{($\arcsec$ yr$^{-1}$)} &
\colhead{(deg)} & \colhead{(km s$^{-1}$)} & \colhead{Notes}
}
\startdata
2MASS J22443167$+$2043433 & 19971005 & L7.5/L6.5 & 2.43 & very red &
21$\pm$5 & 0.33$\pm$0.01 & 130.3$\pm$2.1 & 33$\pm$8  & 1, 2 \\
SDSS J115553.85$+$055957.5 & 20010416 & L7.5        & 1.54 & normal   &
18$\pm$3 & 0.41$\pm$0.03  & 266.3$\pm$3.1 & 35$\pm$6  & 3, 4 \\
SDSS J085758.44$+$570851.4 & 20000406 & L8/L8     & 1.86 & red      &
11$\pm$2 & 0.54$\pm$0.01  & 229.8$\pm$1.4 & 28$\pm$5  & 4, 5\\
SDSS J133148.88$-$011652.5 & 20010524 & L8/L6     & 1.25 & blue     &
18$\pm$4 & 1.11$\pm$0.01 & 201.6$\pm$0.9 & 92$\pm$19 & 4, 6 \\
2MASS J09083803$+$5032088 & 19990306  & L9/L5     & 1.51 & normal   &
10$\pm$2 & 0.62$\pm$0.02 & 219.6 $\pm$2.0 & 30$\pm$5  & 7 \\
SDSS J080531.83$+$481233.1 & 20000425 & L9.5/L4 & 1.10 & blue    &
13$\pm$2 & 0.46$\pm$0.02 & 276.8$\pm$2.1 & 28$\pm$4 & 6, 8 \\
SDSS J120747.17$+$024424.8 & 20000504 & T0/L8     & 1.22 & normal   &
18$\pm$3 & 0.52$\pm$0.02 & 285.5$\pm$2.1 & 43$\pm$8  & 4, 6 \\
SDSS J152039.82$+$354619.8 & 20030326 & T0          & 1.45 & normal   &
18$\pm$3 & 0.48$\pm$0.04  & 144.1$\pm$4.1 & 41$\pm$8  & 4, 9 \\
SDSS J151643.00$+$305344.3 & 20030623 & T0.5        & 1.67 & red      &
31$\pm$6 & 0.13$\pm$0.03 & 271.7$\pm$10.8 & 19$\pm$6  & 9 \\
SDSS J105213.50$+$442255.6AB & 20021213 & T0.5       & 1.43 & normal   &
30$\pm$6\tablenotemark{g} &  0.15$\pm$0.02  & 173.2$\pm$6.4 & 21$\pm$5  & 9, 10 \\
SDSS J075840.32$+$324723.3 & 20011219  & T2          & 0.91 & normal   &
15$\pm$3 & 0.40$\pm$0.02 & 208.7$\pm$2.9 & 28$\pm$6  & 3, 4 \\
2MASS J22541892$+$3123498  & 19980622 & T4          & $-$0.02 & normal &
17$\pm$3 & 0.21$\pm$0.01 & 18.7 $\pm$ 3.9 & 17$\pm$3 &  11 \\
SDSS J000013.54$+$255418.6 & 20030929 & T4.5        & $-$0.09 & normal &
13$\pm$2 & 0.13$\pm$0.02 & 2.6$\pm$5.8 & 8$\pm$2  & 3 \\
SDSS J111009.99$+$011613.0 & 20000505 & T5.5        & 0.07 & very red &
20$\pm$4 & 0.34$\pm$0.02 & 225.6 $\pm$3.3 & 32$\pm$6  & 5 \\
\enddata
\tablenotetext{a}{Coordinates are sexagesimal with the following format: 2MASS Jhhmmssss$+/-$ddmmsss and SDSS Jhhmmss.ss$+/-$ddmmss.s.  These coordinates are valid for the given epoch. }
\tablenotetext{b}{Uncertainty in spectral type is 0.5-1.0 subclasses except for the infrared types of 2MASS 2244$+$20, SDSS 1331$-$01 and SDSS 0805$+$48 which are uncertain by 2.0, 2.5, and 1.5 subclasses respectively.}
\tablenotetext{c}{Uncertainty in $J$--$K$ for the sample is 0.04 mag, except for SDSS 0000+25 which is 0.06 mag, and for SDSS 1110+01 which is 0.07 mag.}
\tablenotetext{d}{Near-infrared color with respect to the typical color for
the type as found from Knapp et al. (2004).}
\tablenotetext{e}{Estimated photometric distance using the magnitude versus spectral type
relationship of Liu et al.\ (2006) which excludes known binaries but not possible binaries.}
\tablenotetext{f}{Proper motions  from Jameson et al.\ (2008) and Faherty et al.\ (2009).}
\tablenotetext{g}{Distance calculation for SDSS 1052+44AB accounts for binarity and assumes equal luminosity for both components.}
\tablecomments{ (1) Discovery by Dahn et al.\ (2002); (2) optical data from
J.~D.\ Kirkpatrick, priv.\ comm.; 
(3) discovery by Knapp et al.\ (2004); (4) SDSS spectrum from Adelman-McCarthy et al.\ (2008); 
(5) discovery by Geballe et al.\ (2002); 
(6) discovery by Hawley et al.\ (2002); (7) discovery and optical data from Cruz
et al.\ (2003); (8) possible L+T binary, Burgasser (2007a); 
(9) discovery by  Chiu et al.\
(2006); (10) resolved L+T binary, Liu et al., in preparation; (11) discovery by
Burgasser et al.\ (2002a).}
\end{deluxetable}
\clearpage

\begin{deluxetable}{lccccccccc}
\rotate
\tabletypesize{\footnotesize}
\tablecaption{Supplementary dwarfs from Cushing et al.\ (2008)}
\tablewidth{600pt}
\tablehead{
\colhead{} & \colhead{Epoch\tablenotemark{a}} & \colhead{Sp.\ Type\tablenotemark{b}} & \colhead{$J$--$K$\tablenotemark{c}} &
\colhead{Relative} & \colhead{D\tablenotemark{e}} &\colhead{$\mu$\tablenotemark{e}} & \colhead{PA\tablenotemark{e}} & \colhead{$v_{\rm tan}$\tablenotemark{e}} &\colhead{}  \\
\colhead{Name\tablenotemark{a}} & \colhead{YYYYMMDD}   & \colhead{(IR/Opt)} & \colhead{(MKO)}   &   \colhead{color\tablenotemark{d}}      &
\colhead{(pc)} & \colhead{($\arcsec$ yr$^{-1}$)} & \colhead{(deg)} & \colhead{(km s$^{-1}$)} & \colhead{Notes}
}
\startdata
2MASS J22244381$-$0158521 & 19981007 & L3.5/L4.5 &  1.92 & red & 11.5$\pm$0.1 & 0.983$\pm$0.001 & 151.6$\pm$0.1 &  53.5$\pm$0.5 & 1, 2 \\
2MASS J00361617$+$1821104 & 20001124 & L4/L3.5   &  1.26 & blue & 8.8$\pm$0.1 & 0.907$\pm$0.001 & 82.4$\pm$0.1 & 37.6$\pm$0.3 & 3, 4 \\
2MASS J15074769$-$1627386 & 19980503 & L5.5/L5   &  1.41 & blue & 7.3$\pm$0.03 & 0.903$\pm$0.001 & 190.1$\pm$0.1 & 31.2$\pm$0.1 & 3, 5 \\
2MASS J08251968$+$2115521 & 19980109 & L6/L7.5   &  1.96 & red & 10.6$\pm$0.1 & 0.585$\pm$0.001 & 240.0$\pm$0.1 & 29.4$\pm$0.3 & 1, 6  \\
DENIS-P J025503.3$-$470049 & 19961029 & L9/L8     &  1.54 & normal & 5.0$\pm$0.1 & 1.149$\pm$0.002 & 119.5$\pm$0.2 &  27.0$\pm$0.5 & 7, 8  \\
SDSS J125453.90$-$012247.5 & 20020212 & T2 &  0.82 & normal & 13.5$\pm$0.3 & 0.490$\pm$0.002 & 284.7$\pm$0.1 & 31.4$\pm$0.7 & 9, 10 \\
2MASS J05591914$-$1404488 & 19981215 & T4.5 &  $-0.16$ & normal & 10.3$\pm$0.1 & 0.660$\pm$0.001 & 121.4$\pm$0.1 &  32.3$\pm$0.3 & 11, 12 \\
\enddata
\tablenotetext{a}{Coordinates are sexagesimal with the following format: 2MASS Jhhmmssss$+/-$ddmmsss, DENIS-P Jhhmmss.s$+/-$hhmmss, and SDSS Jhhmmss.ss$+/-$ddmmss.s.  These coordinates are valid for the given epoch.}
\tablenotetext{b}{Uncertainty in spectral type is 0.5-1.0 subclasses.}
\tablenotetext{c}{Uncertainty in $J$--$K$ for the sample is 0.04 mag, except for DENIS-P J0255-47 for which it is 0.08 mag.}
\tablenotetext{d}{Near-infrared color with respect to the typical color for
the type as found in Knapp et al. (2004).}
\tablenotetext{e}{Trigonometric parallax distances, proper motions, position angles, and tangential velocities taken or derived from values published in Dahn et al.\ (2002), Tinney et al.\  (2003), Vrba et al.\ (2004), and Costa et al.\ (2006).}
\tablecomments{  (1) Discovery by Kirkpatrick et al. (2000); (2) $v$ sin$i$ has been measured at 31
and 32 km s$^{-1}$ (Zapatero Osorio et al. 2006, Reiners \& Basri 2008); (3) discovery by Reid et al. (2000);
(4) $v$ sin$i$ has been measured at 36 and 45 km s$^{-1}$ (Zapatero Osorio et al. 2006, Reiners \& Basri 2008);
(5) $v$ sin$i$ has been measured at 32 km s$^{-1}$ (Reiners \& Basri 2008);
(6) $v$ sin$i$ has been measured at 19 km s$^{-1}$ (Reiners \& Basri 2008);
(7) discovery by Martin et al. 1999; 
(8) $v$ sin$i$ has been measured at 41 and 67 km s$^{-1}$ (Zapatero Osorio et al. 2006, Reiners \& Basri 2008);
(9) discovery by Leggett et al. 2000; 
(10) $v$ sin$i$ has been measured at 28 km s$^{-1}$  (Zapatero Osorio et al. 2006);
(11) discovery by Burgasser et al. (2000);
(12) $v$ sin$i$ has been measured at 23 km s$^{-1}$  (Zapatero Osorio et al. 2006).}
\end{deluxetable}
\clearpage

\begin{deluxetable}{lcccccc}
\tablecaption{Log of observations}
\tabletypesize{\footnotesize}
\tablewidth{500pt}
\tablehead{
\colhead{} & \multicolumn{3}{c}{$Spitzer$ IRS} & \colhead{\hspace*{0.1in}} &
\multicolumn{2}{c}{Gemini NIRI} \\
\cline{2-4}\cline{6-7}
\colhead{} & \multicolumn{2}{c}{Exp. Time (hr)} &
\colhead{Date Obs.} & & \colhead{Exp. Time}& \colhead{Date Obs.}\\
\colhead{Name} & \colhead{1st Order} &
\colhead{2nd Order} &
\colhead{(DD/MM/YY)} & & (hr) & \colhead{(DD/MM/YY)}
}
\startdata
SDSS 0000$+$25 &  2.4 &  3.3 & 18/12/05, 25/01/06 & & & \\
SDSS 0758$+$32 &  0.3  & 0.5 & 24/10/04 & & 3.7 &
08,09,11/12/06 \\
SDSS 0805$+$48 &  0.3   & 1.2 & 23/10/04 & & & \\
SDSS 0857$+$57 &  0.1  &  0.1 &  23/10/04 & & 1.7 & 08/12/06\\
2MASS 0908$+$50 &  0.1 &   0.3  & 23/10/04 & & 1.5 & 22/05/05\\
SDSS 1052$+$44AB &  1.2 &  3.3 &  23/05/06 & & & \\
SDSS 1110$+$01 &  2.4 &  4.8 & 15/01/05, 23/05/05 & & & \\
SDSS 1155$+$05 &  0.5 &  1.6 & 26/01/06 & & & \\
SDSS 1207$+$02 &  1.2  &  2.4 & 03,06/06/05 & & 2.7 & 21/04/05,
22/05/05\\
SDSS 1331$-$01 &  1.2 &  3.3 & 28/01/06 & & & \\
SDSS 1516$+$30 &  1.3 &  3.3 &  15/08/05 & & & \\
SDSS 1520$+$35 &  0.5  & 1.6 & 15/08/05 & & & \\
2MASS 2244$+$20 &  0.3  & 0.8  & 10/12/04 & & 2.1 & 20/06/05\\
2MASS 2254$+$31  & 1.5 &  4.8 & 05/01/05 & & & \\
\enddata
\end{deluxetable}
\clearpage

\begin{deluxetable}{lllllllr}
\rotate
\tablecaption{Adopted (and range of) model parameters}
\tabletypesize{\footnotesize}
\tablewidth{600pt}
\tablehead{
\colhead{} &\colhead{Sp.~Type} & \colhead{\teff} & \colhead{\logg} & \colhead{\fsed} & \colhead{$K_{zz}$} & Radius & \colhead{} \\
\colhead{Name} & \colhead{(IR/Opt)} & \colhead{(K)} & \colhead{} & \colhead{(nc$=$``no cloud'')} & \colhead{(cm$^2$ s$^{-1}$)} &\colhead{R$_{\odot}$} & \colhead{Notes\tablenotemark{a}} \\
}
\startdata
2MASS 2224$-$01 & L3.5/L4.5 & 1400 (1300 -- 1500) & 4.5 (4.5 -- 5.0) & 1 (1 -- 2) & $10^2$ (0 -- $10^6$) & 0.11 -- 0.13 & 1, 2 \\
2MASS 0036$+$18 & L4/L3.5 & 1800 (1700 -- 1800)  & 5.0 (5.0) & 3 (2 -- 3) & $10^2$ (0 -- $10^6$)  &  0.11  & 3 \\
2MASS 1507$-$16 & L5.5/L5 & 1600 (1600 -- 1700) & 5.5 (5.5) & 3 (2 -- 3) & $10^6$ (0 -- $10^6$)  & 0.08 & 4 \\
2MASS 0825$+$21 & L6/L7.5 & 1200 (1100 -- 1300) & 5.5 (5.0 -- 5.5) & 2 (2) & $10^4$  ($10^4$ -- $10^6$) & 0.08 -- 0.10 & 5 \\
2MASS 2244$+$20 & L7.5/L6.5 & 1100 (1000 -- 1200) & 5.0 (4.5 -- 5.0) & 1 (1) & $10^6$  ($10^2$ -- $10^6$) & 0.10 -- 0.13 & 6 \\
SDSS 1155$+$05  & L7.5 & 1400 (1300 -- 1500) & 5.0 (4.5 -- 5.0) & 2 (2) & $10^4$  ($10^2$ -- $10^6$) & 0.10 -- 0.13 &  \\
SDSS 0857$+$57  & L8/L8 & 1200 (1100 -- 1300) & 4.5 (4.5 -- 5.0) &  2 (1 -- 2) & $10^6$  ($10^2$ -- $10^6$) & 0.10 -- 0.13 & 7 \\
SDSS 1331$-$01  & L8/L6 & 1600 (1500 -- 1600) & 5.0 (5.0 -- 5.5) & 3 (3 -- 4) & $10^4$  (0 -- $10^6$) & 0.08 -- 0.11 &  \\
2MASS 0908$+$50 & L9/L5 & 1300 (1300 -- 1400) & 5.5 (5.0 -- 5.5) & 3 (2 -- 3) & $10^6$  ($10^2$ -- $10^6$) & 0.08 -- 0.11 & 8 \\
DENIS 0255$-$47 & L9/L8 & 1300 (1200 -- 1300) & 5.0 (5.0 -- 5.5) & 2 (2) & $10^4$  ($10^2$ -- $10^6$) & 0.08 -- 0.10 & 9 \\
SDSS 0805$+$48  & L9.5/L4 & 1600 (1600) & 5.5 (5.5) & 3 (2 -- 3) & $10^4$  ($10^2$ -- $10^6$) & 0.08 & 10 \\
SDSS 1207$+$02  & T0/L8 & 1300 (1200 -- 1400) & 5.0 (4.5 -- 5.5) & 3 (2 -- 3) & $10^4$  ($10^2$ -- $10^6$) & 0.08 -- 0.13 & 11\\
SDSS 1520$+$35  & T0 & 1400 (1400) & 5.0 (4.5 -- 5.0) & 2 (2) & $10^2$ (0 -- $10^2$) & 0.11 -- 0.13 & 12\\
SDSS 1516$+$30  & T0.5 & 1100 (1000 -- 1100) & 4.5 (4.5) & 2 (1 -- 2) & $10^4$  ($10^2$ -- $10^6$) & 0.12 & 13 \\
SDSS 1052$+$44AB & T0.5 & 1300 (1200 -- 1400) & 5.5 (5.0 -- 5.5) & 3 (2 -- 3) & $10^4$  (0 -- $10^6$) & 0.08 -- 0.11 & 14 \\
SDSS 0758$+$32  & T2 & 1100 (1100 -- 1200) & 4.5 (4.5) & 3 (3 -- 4) & $10^4$  ($10^4$ -- $10^6$) & 0.12 -- 0.13 & \\
SDSS 1254$-$01  & T2 & 1200 (1100 -- 1300) & 5.0 (4.5-- 5.0) & 4 (3 -- 4) & $10^4$  ($10^2$ -- $10^6$) & 0.10 -- 0.12 & \\
2MASS 2254$+$31 & T4 & 1200 (1100 -- 1300) & 5.0 (4.5 -- 5.5) & nc (4 -- nc) & $10^2$ (0 -- $10^4$) & 0.08 -- 0.12 & \\
2MASS 0559$-$14 & T4.5 & 1200 (1200 -- 1300) & 4.5 (4.5 -- 5.0) & nc (nc) & $10^2$ (0 -- $10^2$) & 0.10 -- 0.12 & 15 \\
SDSS 0000$+$25  & T4.5 & 1200 (1100 -- 1200) & 5.0 (4.5 -- 5.5) & nc (4 -- nc) & $10^2$ (0 -- $10^4$) & 0.08 -- 0.12 & \\
SDSS 1110$+$01  & T5.5 & 1000 (900 -- 1100) & 4.5 (4.5) & 4 (4 -- nc) &  $10^4$  (0 -- $10^6$) & 0.11 -- 0.12 & \\
\enddata
\tablenotetext{a}{Notes:\\
(1) Model fit is poor for this very red L dwarf and the estimated uncertainty in \teff\ is $\pm$ 200K.  \\
(2) \teff\ / \logg\ / \fsed\ / $K_{zz}$ fits excluded by radius test: 1300-1400/5.5/1-2/$10^2$-$10^6$, 1600-1700/4.5-5.0/1-2/0-$10^6$.\\
(3) \teff\ / \logg\ / \fsed\ / $K_{zz}$ fits excluded by radius test: 1600-1700/5.5/2-3/0-$10^6$, 1700-1800/4.5/1-3/0-$10^6$, 1900/4.5-5.5/1-4/0.\\
(4) \teff\ / \logg\ / \fsed\ / $K_{zz}$ fits excluded by radius test: 1400-1500/5.5/3/$10^4$-$10^6$, 1600-1700/5.0/2-3/0-$10^6$.\\
(5) \teff\ / \logg\ / \fsed\ / $K_{zz}$ fits excluded by radius test: 1100-1400/4.5/1-2/0-$10^6$, 1100-1200/5.5/2/$10^4$-$10^6$.\\
(6) \teff\ / \logg\ / \fsed\ / $K_{zz}$ fits excluded by radius test: 900/4.5-5.5/1/$10^2$-$10^6$, 1200/5.5/1/$10^4$-$10^6$.\\
(7) \teff\ / \logg\ / \fsed\ / $K_{zz}$ fits excluded by radius test: 1100/5.5/2/$10^6$.\\
(8) \teff\ / \logg\ / \fsed\ / $K_{zz}$ fits excluded by radius test: 1400-1500/4.5/2/$10^2$-$10^6$.\\
(9) \teff\ / \logg\ / \fsed\ / $K_{zz}$ fits excluded by radius test: 1200/4.5/2/$10^4$, 1400/4.5-5.0/2/0-$10^6$.\\
(10) \teff\ / \logg\ / \fsed\ / $K_{zz}$ fits excluded by radius test: 1600/5.0/3/$10^6$, 1700/4.5-5.0/1-3/0-$10^6$.\\
(11) \teff\ / \logg\ / \fsed\ / $K_{zz}$ fits excluded by radius test: 1400/4.5/2/$10^2$-$10^6$. \\
(12) \teff\ / \logg\ / \fsed\ / $K_{zz}$ fits excluded by radius test: 1500/4.5/2/0-$10^2$. \\
(13) \teff\ / \logg\ / \fsed\ / $K_{zz}$ fits excluded by radius test: 900/4.5-5.5/1-2/$10^2$-$10^6$, 1000/5.0-5.5/2/$10^4$-$10^6$. \\
(14) Known to be an L$+$T binary (Liu et al. in preparation); this fit assumes the binary is composed of an identical pair of dwarfs.\\
(15) \teff\ / \logg\ / \fsed\ / $K_{zz}$ fits excluded by radius test: 1100/4.5-5.5/4-no cloud/0-$10^6$, 1200-1300/5.5/4-no cloud/0-$10^6$.\\
}
\end{deluxetable}
\clearpage


\begin{deluxetable}{lcccccccc}
\rotate
\tablecaption{Parameters of known or possible binaries}
\tabletypesize{\footnotesize}
\tablewidth{550pt}
\tablehead{
\colhead{} & \colhead{Comb. Sp. Type} & \multicolumn{2}{c}{\teff/\logg/\fsed/${\rm log}(K_{zz})$} &
\colhead{$D$} & \colhead{$v_{tan}$} & \colhead{Age} & \multicolumn{2}{c}{Mass ($M_{Jup}$)}\\
\cline{3-4}\cline{8-9}
\colhead{Name}     & \colhead{(IR/Opt)} & \colhead{Primary}   & \colhead{Secondary}  &
\colhead{(pc)}  & \colhead{(km/s)} & \colhead{(Gyr)} & \colhead{Pri.} &
\colhead{Sec.}
}
\startdata
2MASS 0036$+$18 & L4/L3.5 & 1650/5.5/2.5/1 & 1650/5.5/2.5/1 & 8.8$\pm$0.1 & 37.6$\pm$0.3 & 1 -- 10 & 50 -- 80 & 50 -- 80 \\
SDSS 0805$+$48 & L9.5/L4 & 1600/5.0/3/4 & 1000/4.5/nc/4 & 24$\pm$5 & 52$\pm$11  & 0.2 -- 0.4 & 30 -- 40 & 15 -- 25 \\
SDSS 1052$+$44AB & T0.5      & 1400/5.0/2/4 & 1100/5.0/4/4 & 30$\pm$6 & 21$\pm$5  & 0.5 -- 1 & 40 -- 50 & 30 -- 40 \\
2MASS 0559$-$14  & T4.5 & 1200/5.5/4/4 &  1200/5.5/4/4 & 10.3$\pm$0.1 & 32.2$\pm$0.3     & 2 -- 10 & 50 -- 80 & 60 -- 80 \\
\enddata

\end{deluxetable}

\clearpage

\begin{deluxetable}{llcll}
\tablecaption{Estimated masses and ages\tablenotemark{a}}
\tablewidth{350pt}
\tablehead{
\colhead{} & \colhead{Sp.\ Type} & \colhead{Mass} & \colhead{Age} & \colhead{$v_{\rm tan}$} \\
\colhead{Name}     & \colhead{(IR/Opt)} & \colhead{($M_{Jup}$)}    & \colhead{(Gyr)}             & \colhead{(km s$^{-1}$)}
}
\startdata
2MASS 0036$+$18\tablenotemark{b}  & L4/L3.5 & 25 -- 65 & 0.1 -- 1  & 37.6 $\pm$ 0.3 \\
2MASS 1507$-$16                   & L5.5/L5 & 50 -- 70 & 0.4 -- 10  & 31.2 $\pm$ 0.1\\
2MASS 0825$+$21                   & L6/L7.5 & 30 -- 65 & 0.4 -- 10  & 29.4 $\pm$ 0.3\\
2MASS 2244$+$20                   & L7.5/L6.5 & 15 -- 50 & 0.2 -- 2 & 33 $\pm$ 8\\ 
SDSS 1155$+$05                    & L7.5    & 15 -- 55 & 0.1 -- 1  & 35 $\pm$ 6  \\
SDSS 0857$+$57                    & L8/L8   & 15 -- 45 & 0.1 -- 2 & 28 $\pm$ 5  \\
SDSS 1331$-$01                    & L8/L6   & 35 -- 70 & 0.3 -- 10   & 92 $\pm$ 19  \\
2MASS 0908$+$50                   & L9/L5   & 30 -- 65 & 0.3 -- 10   & 30 $\pm$ 5  \\
DENIS 0255$-$47                   & L9/L8   & 25 -- 65 & 0.3 -- 10  & 27.0 $\pm$ 0.5\\
SDSS 0805$+$48\tablenotemark{b}   & L9.5/L4 & 50 -- 70 & 0.5 -- 10  & 28 $\pm$ 4  \\
SDSS 1207$+$02                    & T0/L8   & 15 -- 65 & 0.1 -- 10  & 43 $\pm$ 8  \\
SDSS 1520$+$35                    & T0      & 15 -- 50 & 0.1 -- 1   & 41 $\pm$ 8  \\
SDSS 1516$+$30                    & T0.5    & 10 -- 30 & 0.04 -- 1  & 19$\pm$6 \\
SDSS 0758$+$32                    & T2      & 10 -- 25 & 0.04 -- 0.4  & 28 $\pm$ 6    \\
SDSS 1254$-$01                    & T2      & 10 -- 50 & 0.1 -- 2  & 31.4 $\pm$ 0.7\\
2MASS 2254$+$31                   & T4      & 15 -- 65 & 0.1 -- 10  & 17 $\pm$ 3 \\
2MASS 0559$-$14\tablenotemark{b}  & T4.5    & 10 -- 45 & 0.1 -- 2   & 32.3 $\pm$ 0.3\\
SDSS 0000$+$25                    & T4.5    & 10 -- 65 & 0.1 -- 10 & 8 $\pm$ 2    \\
SDSS 1110$+$01                    & T5.5    & 7 -- 25 & 0.1 -- 1  & 32 $\pm$ 6    \\
\enddata
\tablenotetext{a}{Based on evolutionary models (Saumon \& Marley 2008), using the values of \teff\ and \logg\ given in Table~4 with an assumed $\pm\sigma$(\teff)$=$120~K and $\pm\sigma$(\logg)$=$0.4 dex.}
\tablenotetext{b}{Values for a single dwarf; see Table 5 for binary values.}
\end{deluxetable}

\clearpage

\begin{figure}
\includegraphics[angle=-90,scale=.65]{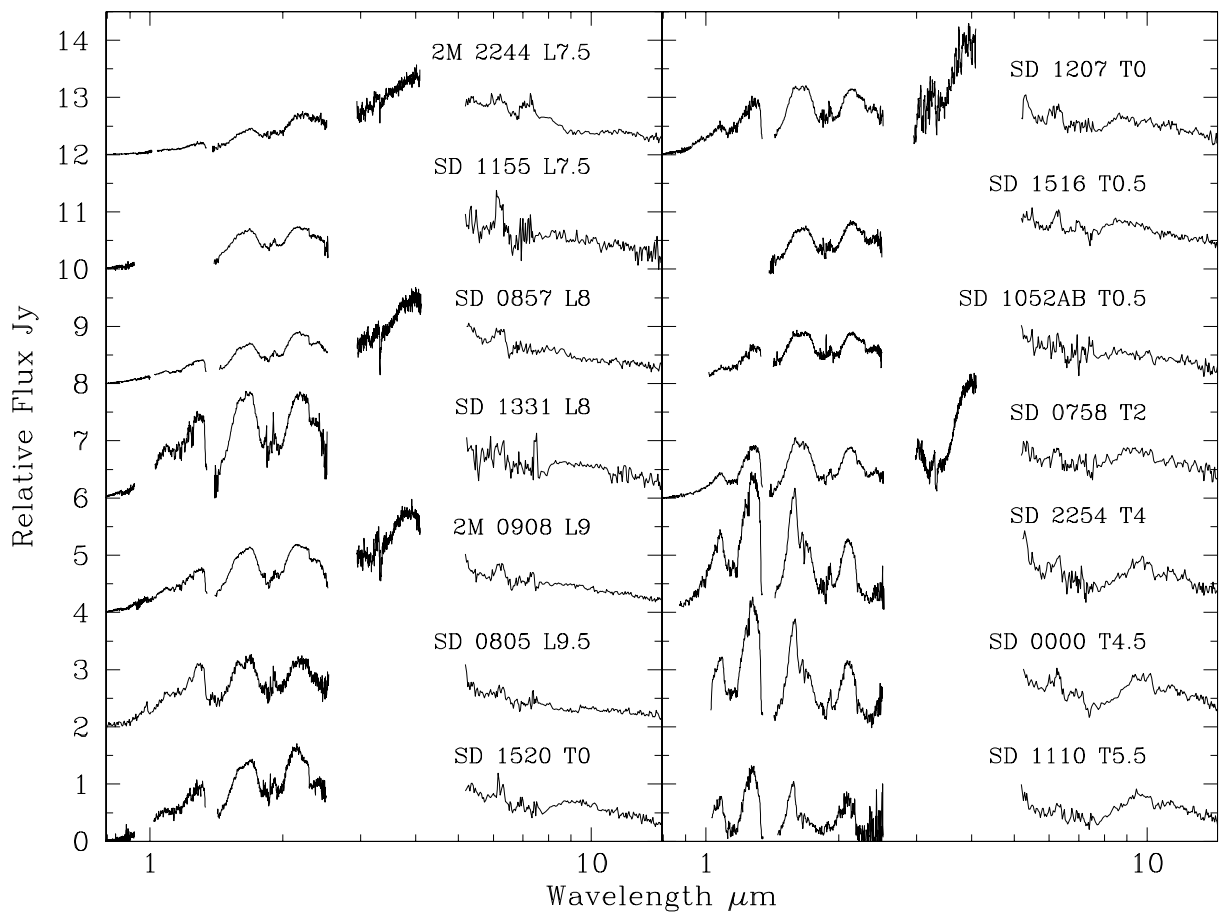}
\caption{Red to mid-infrared spectra for the 14 L and T dwarfs listed in Table~1.  The names of the dwarfs have been abbreviated for clarity and infrared spectral types are given.  The spectra have been normalized to the flux at 5.4~\microns\ and vertically offset in increments of 2 Jy.
\label{fig1}}
\end{figure}

\clearpage

\begin{figure}
\includegraphics[angle=0,scale=.60]{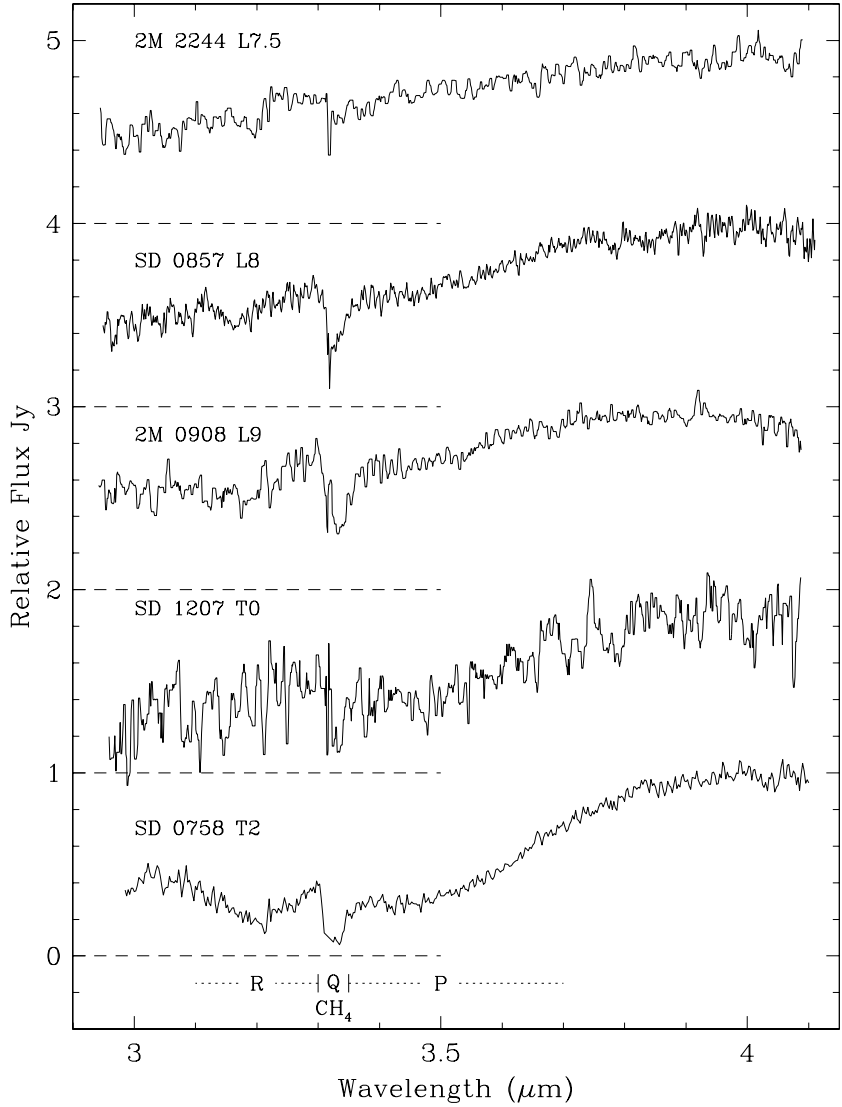}
\caption{3.0 -- 4.1~\microns\ spectra of five L and T dwarfs listed in Table~1.  The names of the dwarfs have been abbreviated for clarity and infrared spectral types are given.  The spectra have been normalized to the flux peak at 4~\microns\ and vertically offset in increments of 1 Jy.  The dashed lines indicate the zero flux level for each dwarf.
\label{fig2}}
\end{figure}

\clearpage

\begin{figure}
\includegraphics[angle=0,scale=1.0]{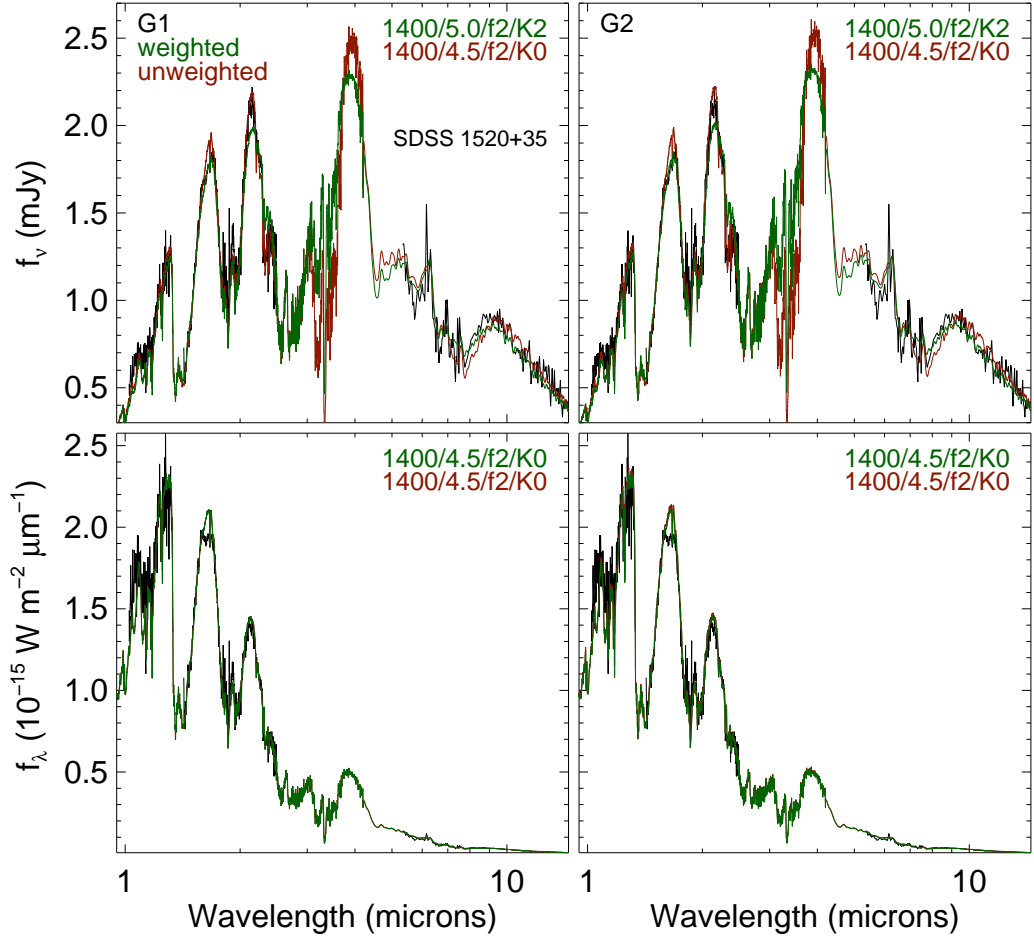}
\caption{Fits to the SED of the T0 dwarf SDSS 1520$+$35 (black lines) using the eight statistics described in \S 5. For each statistic, the best-fit model parameters are shown as \teff\ / \logg\ / \fsed\ / $\log(K_{zz})$ where K0 indicates $K_{zz}$=0, the chemical equilibrium case.  The labeled values of \logg\ = 4.477 and 5.477 are rounded off to 4.5 and 5.5, respectively.  Note in the f$_{\lambda}$ case the best weighted and unweighted solutions are the same.
\label{fig3}}
\end{figure}

\clearpage

\begin{figure}
\includegraphics[angle=0,scale=1.0]{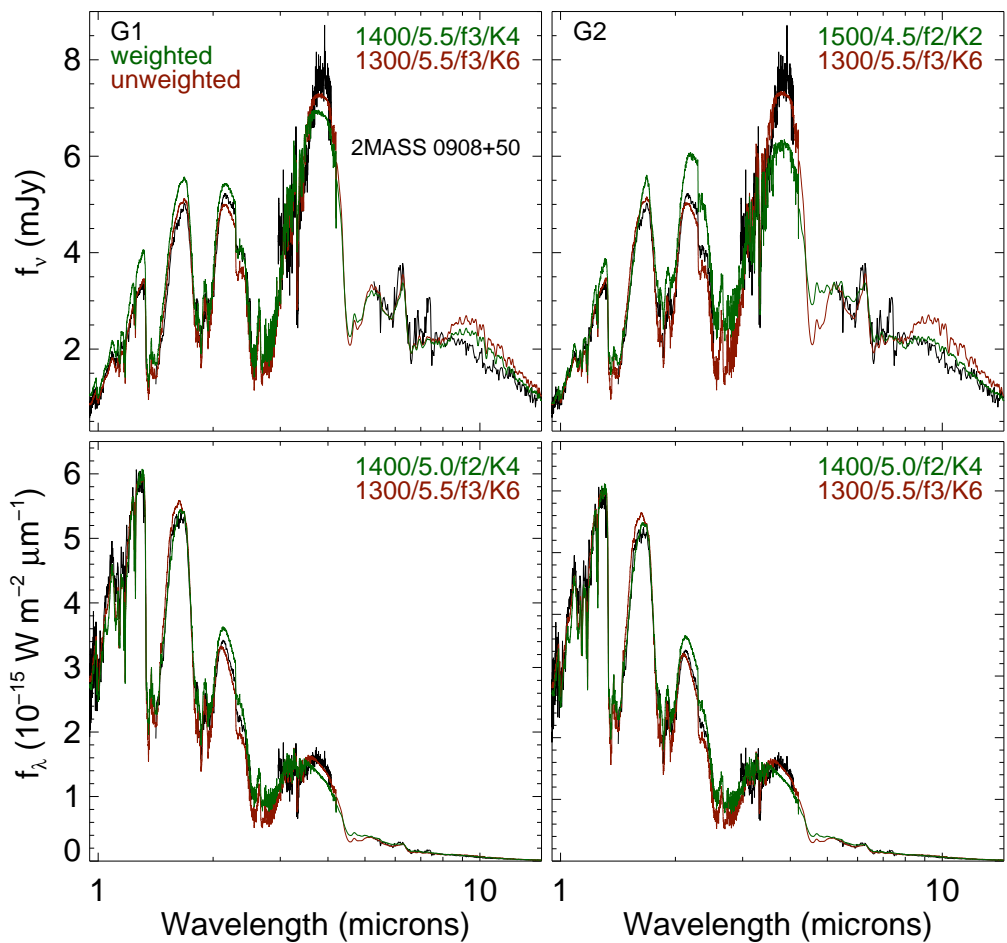}
\caption{Fits to the SED of the L9 dwarf (L5 optically) 2MASS 0908$+$50 (black lines) using the eight statistics described in \S 5. For each statistic, the best-fit model parameters are shown as \teff\ / \logg\ / \fsed\ / $\log(K_{zz})$ as in Figure 3.
\label{fig4}}
\end{figure}

\clearpage

\begin{figure}
\includegraphics[angle=0,scale=.70]{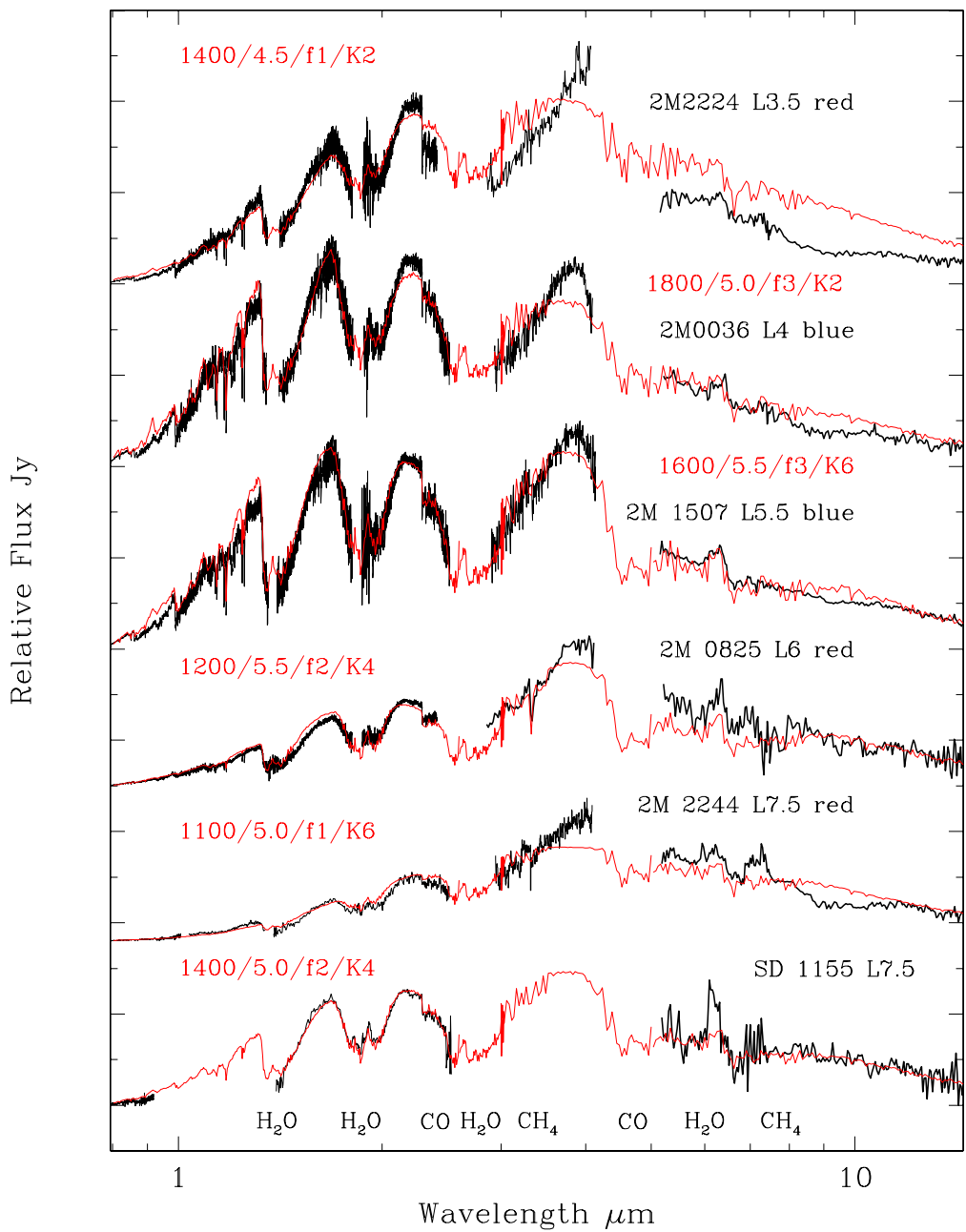}
\caption{Model spectra (red curves) with the adopted parameters from Table 4 are plotted with the observed spectra (black curves) of the L3 -- L7.5 dwarfs in our sample.  The model parameters are labeled as in Figure~3.  The spectra are normalized to the flux at 5.4~\microns\ and vertically offset for clarity.   The model spectra have been smoothed to match the resolution of the observed spectra.
\label{fig5}}
\end{figure}

\clearpage

\begin{figure}
\includegraphics[angle=0,scale=.70]{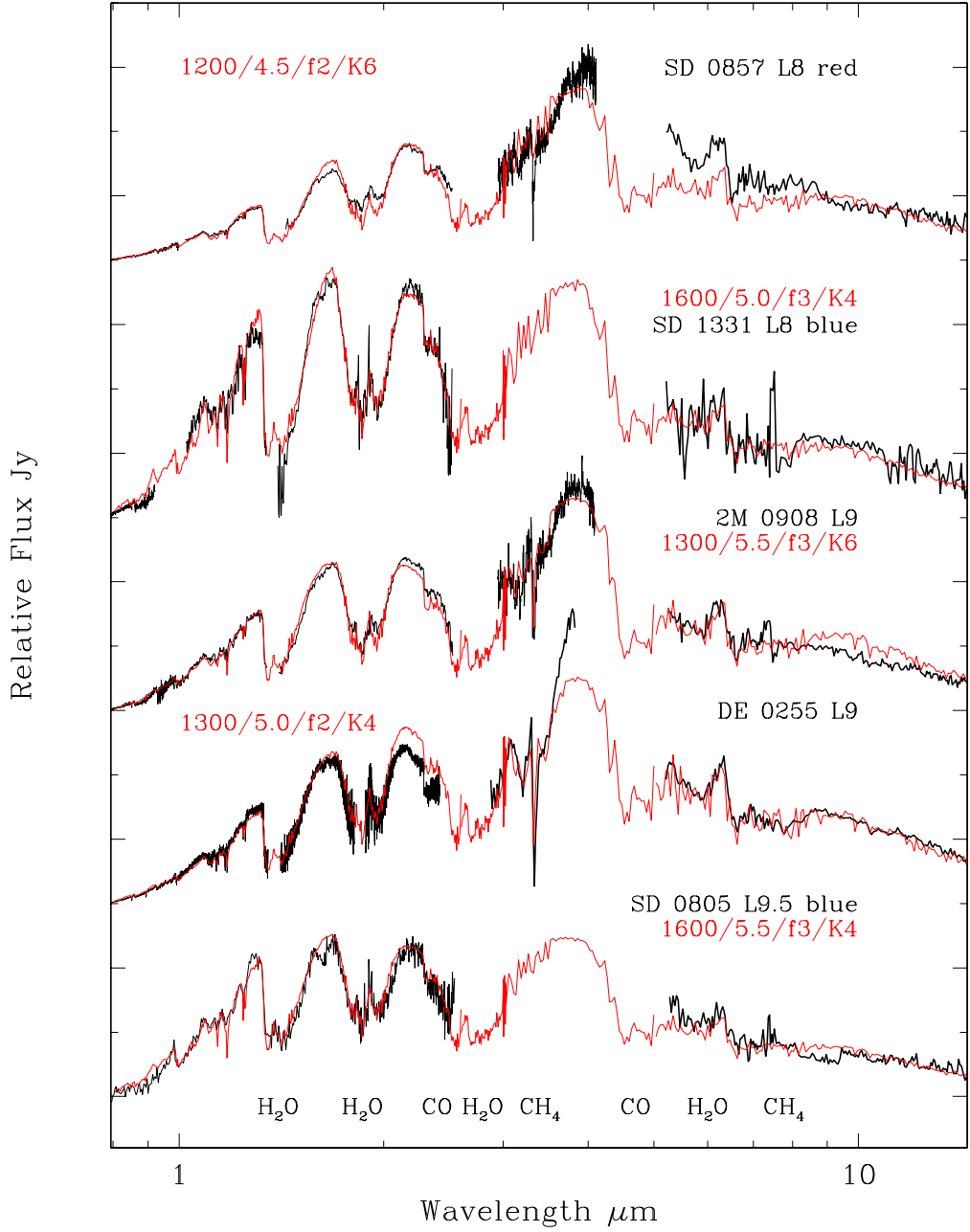}
\caption{Same as Figure~5, but for the L8 -- L9.5 dwarfs in our sample.
\label{fig6}}
\end{figure}

\clearpage

\begin{figure}
\includegraphics[angle=0,scale=.70]{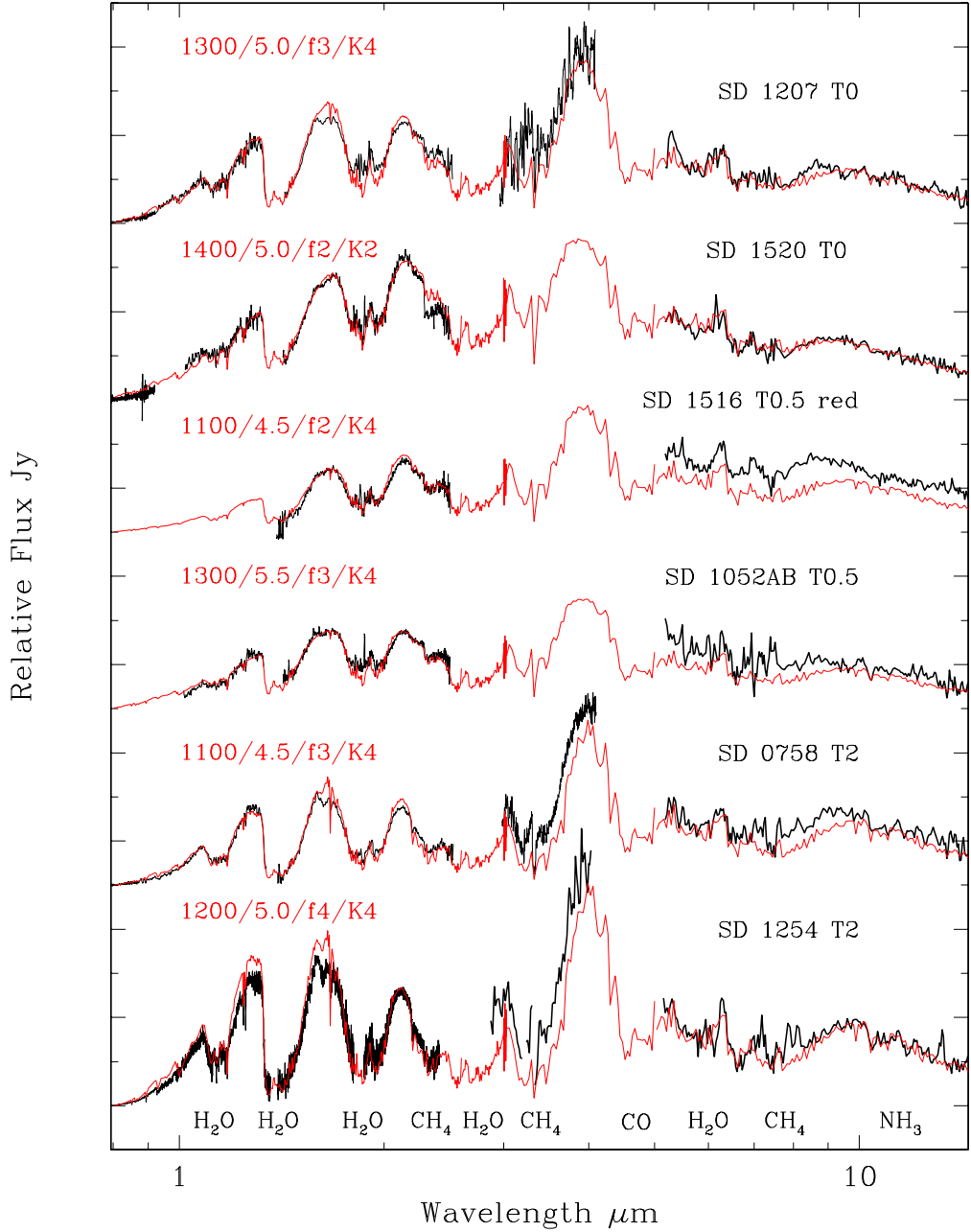}
\caption{Same as Figure~5, but for the early-T dwarfs in our sample.  The known binary SDSS~1052$+$44AB is modeled here as an identical pair of dwarfs.
\label{fig7}}
\end{figure}

\clearpage

\begin{figure}
\includegraphics[angle=0,scale=.70]{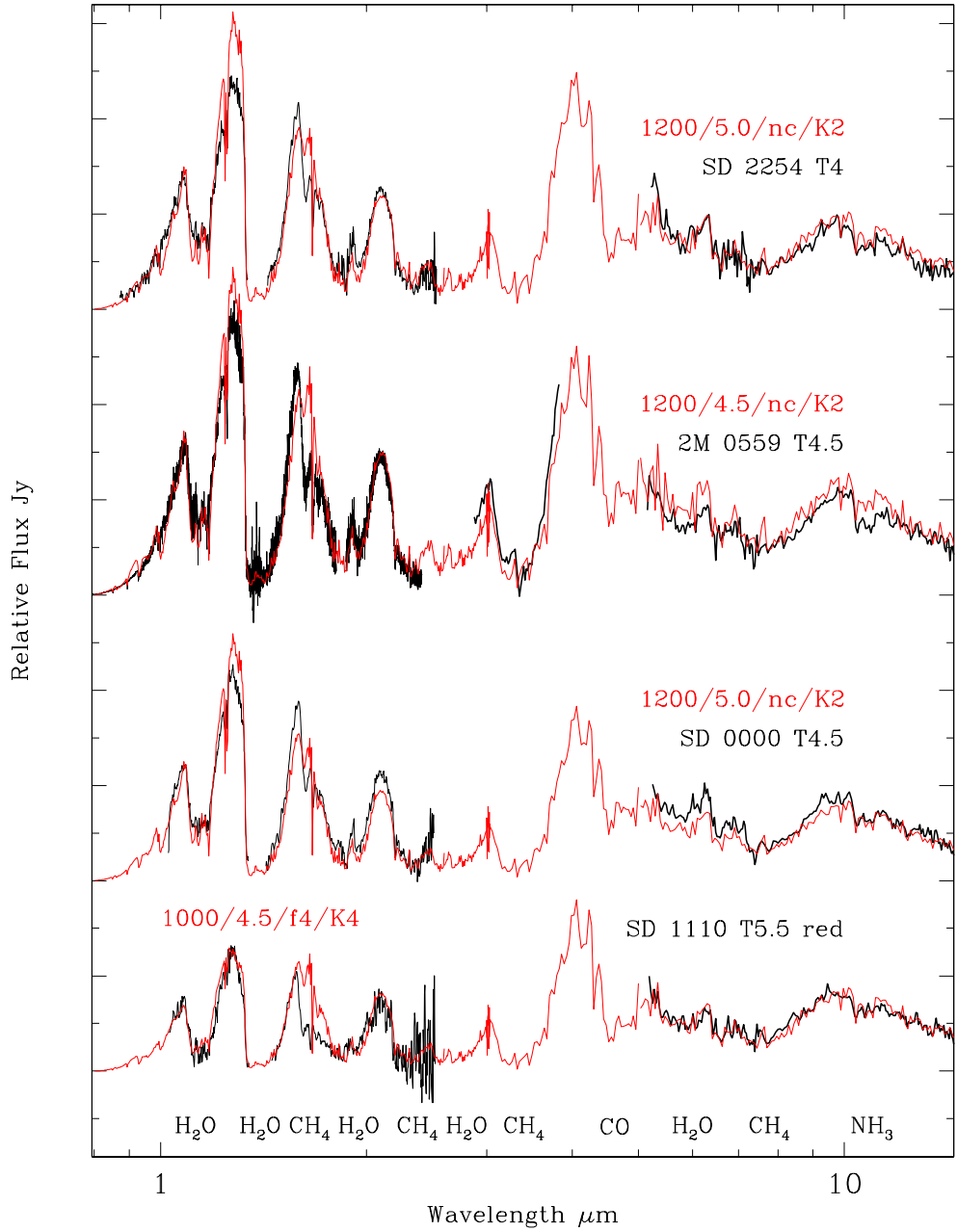}
\caption{Same as Figure~5, but for the mid-T dwarfs in our sample.  The ``nc'' label refers to ``no cloud'' models.
\label{fig8}}
\end{figure}

\clearpage

\begin{figure}
\includegraphics[angle=0,scale=.50]{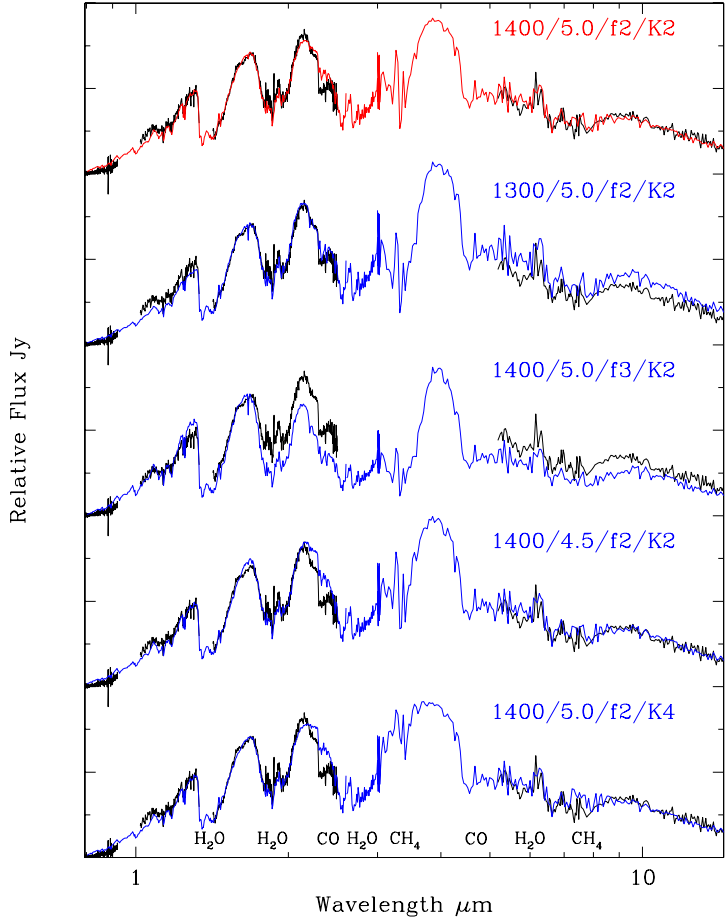}
\caption{Effects on the model fits to the observed spectrum of the T0 dwarf SDSS 1520$+$35 (black curves) as the model parameters are varied as described in \S 5.1. The topmost spectra show the adopted model (red curve) with the indicated values of \teff\ / \logg\ / \fsed\ / $\log(K_{zz})$. The other model spectra (blue curves) demonstrate the effect of changing each parameter in our parametric grid. Changing \teff\ affects the ratio of the near- to mid-infrared fluxes, changing \fsed\ affects the near-infrared flux slope, changing \logg\ affects the $K$-band brightness, and changing $K_{zz}$ affects the strength of the CH$_4$ absorption bands centered at 7.65, 3.3, and more subtly at 2.2~\microns. The spectra are normalized at 5.4~\microns\ and vertically offset for clarity.  The model spectra have been smoothed to match the resolution of the observed spectra.
\label{fig9}}
\end{figure}

\clearpage

\begin{figure}
\includegraphics[angle=0,scale=.70]{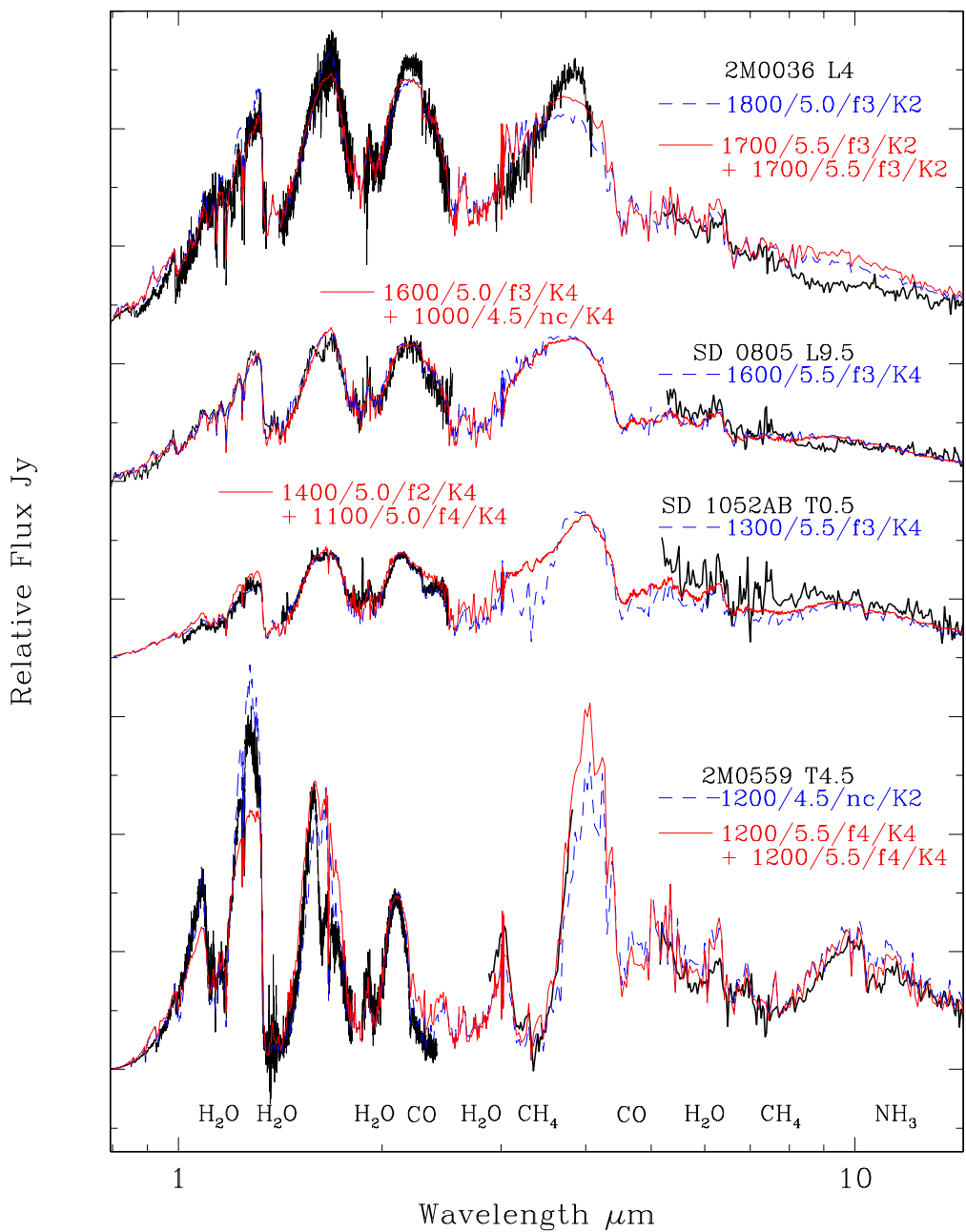}
\caption{Observed spectra of four dwarfs (black lines) that can be modeled as binary systems. The blue curves (dashed lines) are the adopted model fits to each dwarf if single, except for the known binary SDSS~1052$+$44AB which is modeled as an identical pair of dwarfs. The red curves (solid lines) show the modeled composite binary spectra, where we allow the components of SDSS~1052$+$44AB to differ. In all cases the quality of the fits are not significantly different.  The spectra have been normalized to the flux at 5.4 \microns\ and vertically offset for clarity.  Model parameters are labeled as in Figure~3. The model spectra have been smoothed to match the resolution of the observed spectra. 
\label{fig10}}
\end{figure}

\clearpage

\begin{figure}
\includegraphics[angle=-90,scale=.60]{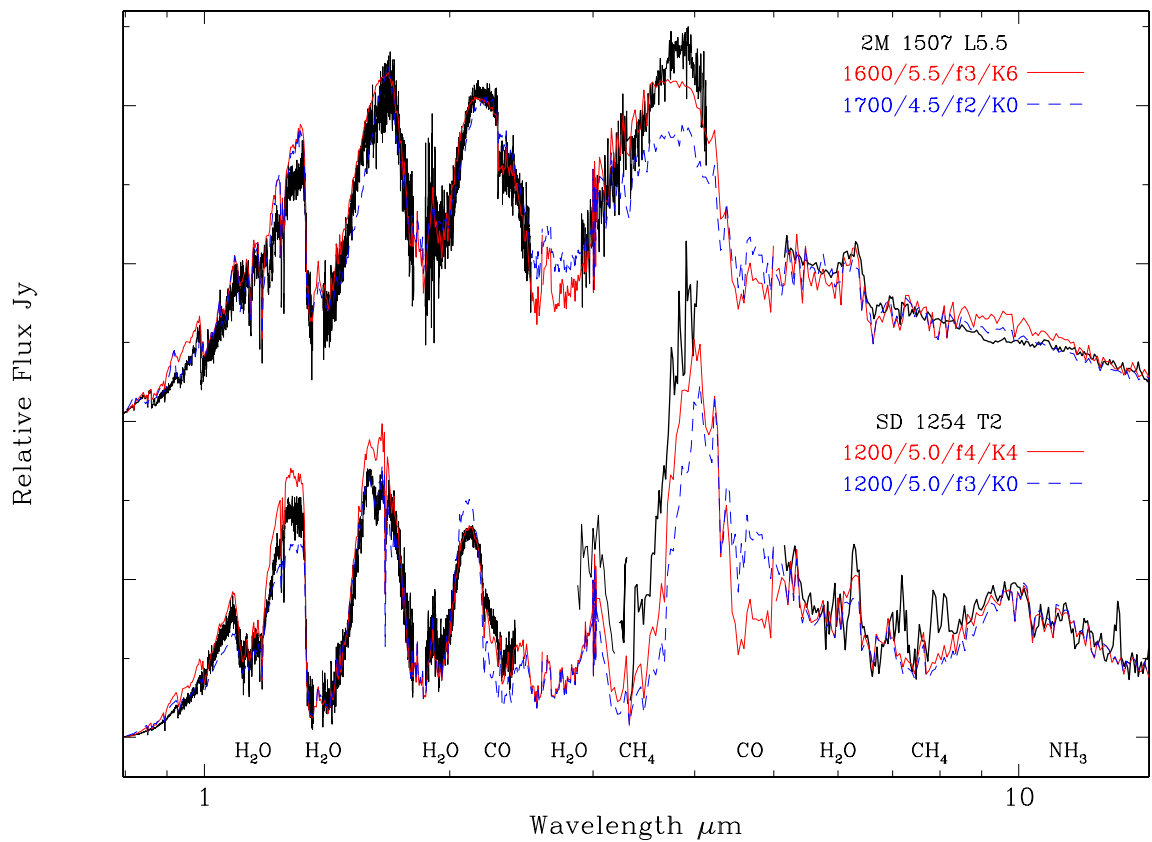}
\caption{Comparison of model fits to mid-L and early-T dwarfs with and without vertical mixing in the atmosphere.  The black curves represent the observed spectra, the red curves (solid lines) show our best fitting models with vertical mixing, and the blue curves (dashed lines) show the model fits obtained by Cushing et al.\ (2008) without vertical mixing. In both cases the models with vertical mixing provide a better fit to the 2 -- 4 $\mu$m region. The spectra have been normalized to the flux at 5.4 \microns\ and vertically offset for clarity.  The model spectra have been smoothed to match the resolution of the observed spectra. 
\label{fig11}}
\end{figure}

\clearpage

\begin{figure}
\includegraphics[angle=-90,scale=.60]{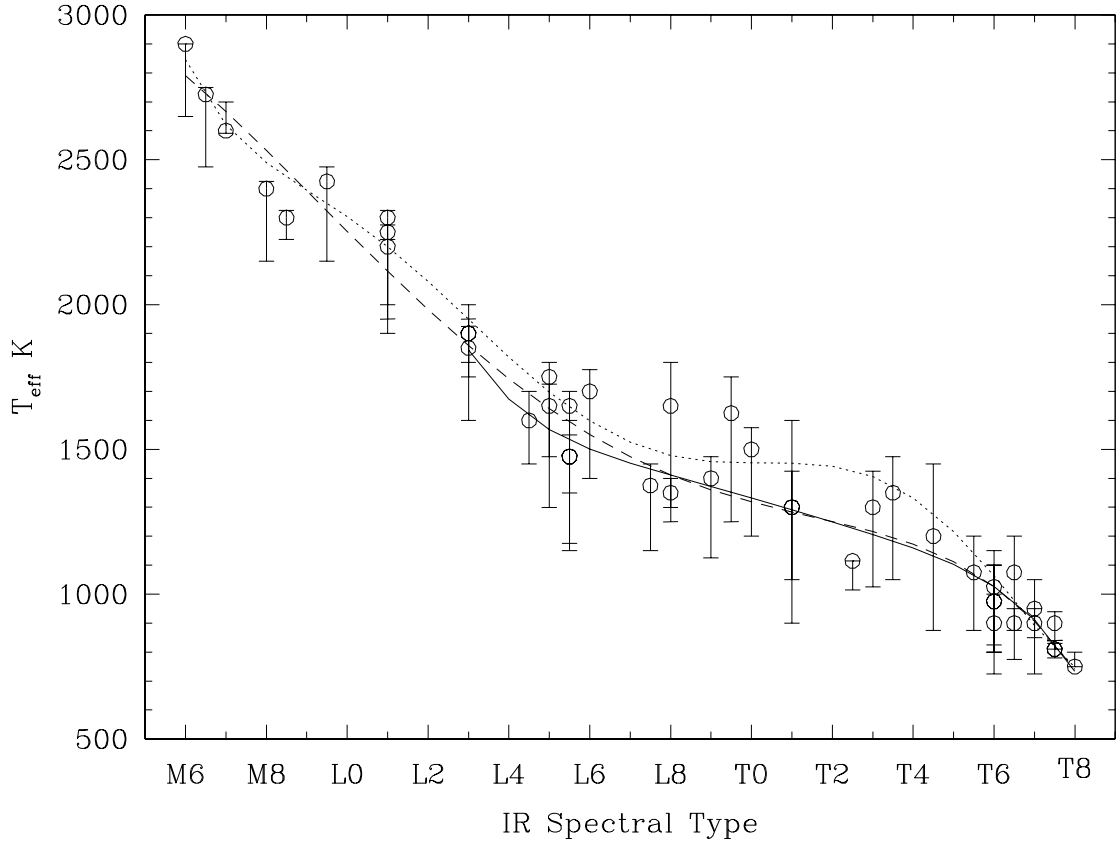}
\caption{Effective temperatures of L and T dwarfs based on measured bolometric luminosities: a revised version of Figure~6 from Golimowski et al.\ (2004a, see text).   The solid line is a fifth order
polynomial fit to the data between spectral types L3 and T8, and the dashed line is a similar fit between spectral types M6 and T8.  The dotted line shows the relationship derived by Golimowski et al..
\label{fig12}}
\end{figure}

\clearpage

\begin{figure}
\includegraphics[angle=0,scale=.60]{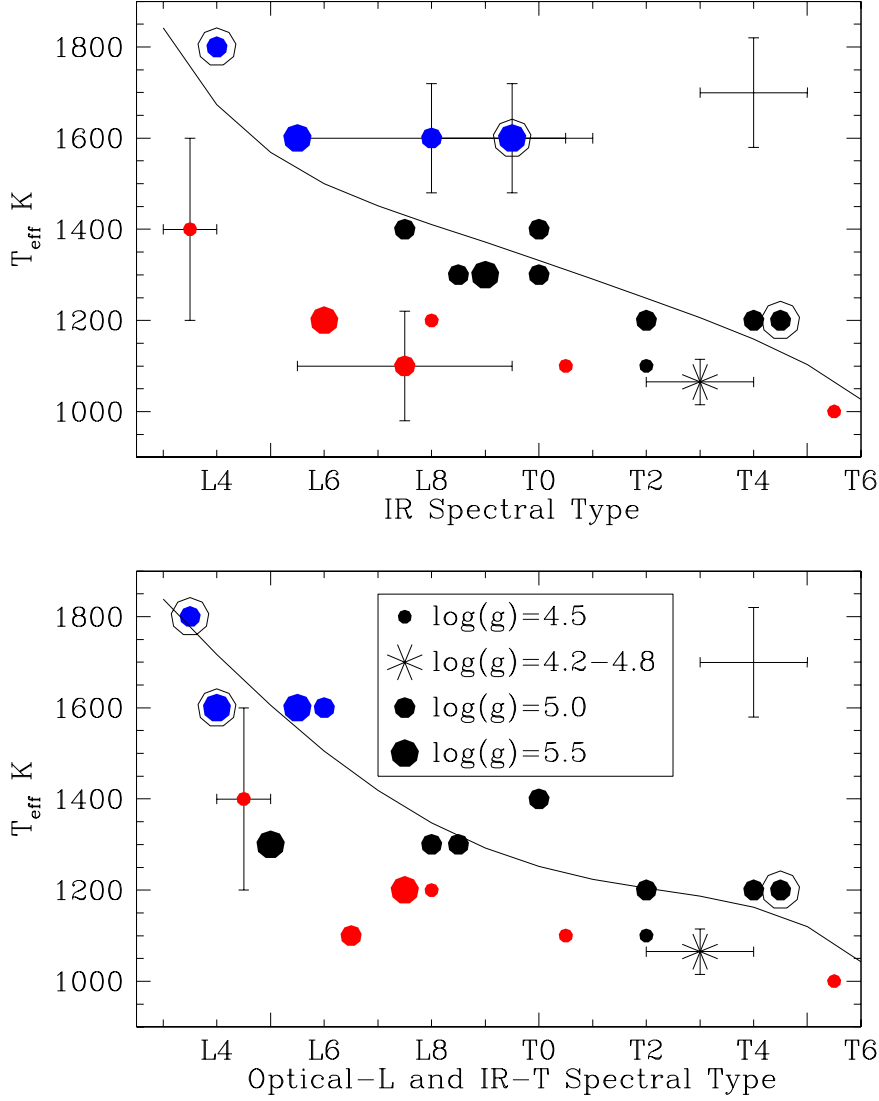}
\caption{Effective temperatures for the sample. The top panel plots temperature as a function of infrared spectral type and also shows the polynomial fit for L3 to T8 types from Figure 12.  The lower panel uses optical types for the L dwarfs and show the polynomial fit found using optical types for the L dwarfs. In order to separate datapoints both the optical and infrared type for DENIS~0255$-$47 have been changed by half a subclass to L8.5.  The blue and red symbols represent dwarfs with unusually blue or red near-infrared colors, respectively.  The sizes of the symbols are proportional to the surface gravity, as indicated in the lower panel.  The ringed symbols represent known or possible binary systems.  The asterisk denotes HN Peg B, which was studied by Leggett et al.\ (2008) with the same model grid used in this paper.  The error bar shows the typical uncertainties in temperature and spectral type. The smaller \teff\  uncertainties for HN Peg B are indicated.  The uncertainty in \teff\ is larger for the very red L dwarf 2MASS~2224$-$01 (L3.5), and the uncertainties in infrared spectral type are larger for the red L dwarf 2MASS 2244$+$20 (L7.5), and the blue L dwarfs SDSS 0805$+$48 (L9.5) and SDSS 1331$-$01 (L8). 
\label{fig13}}
\end{figure}

\clearpage

\begin{figure}
\includegraphics[angle=-90,scale=.70]{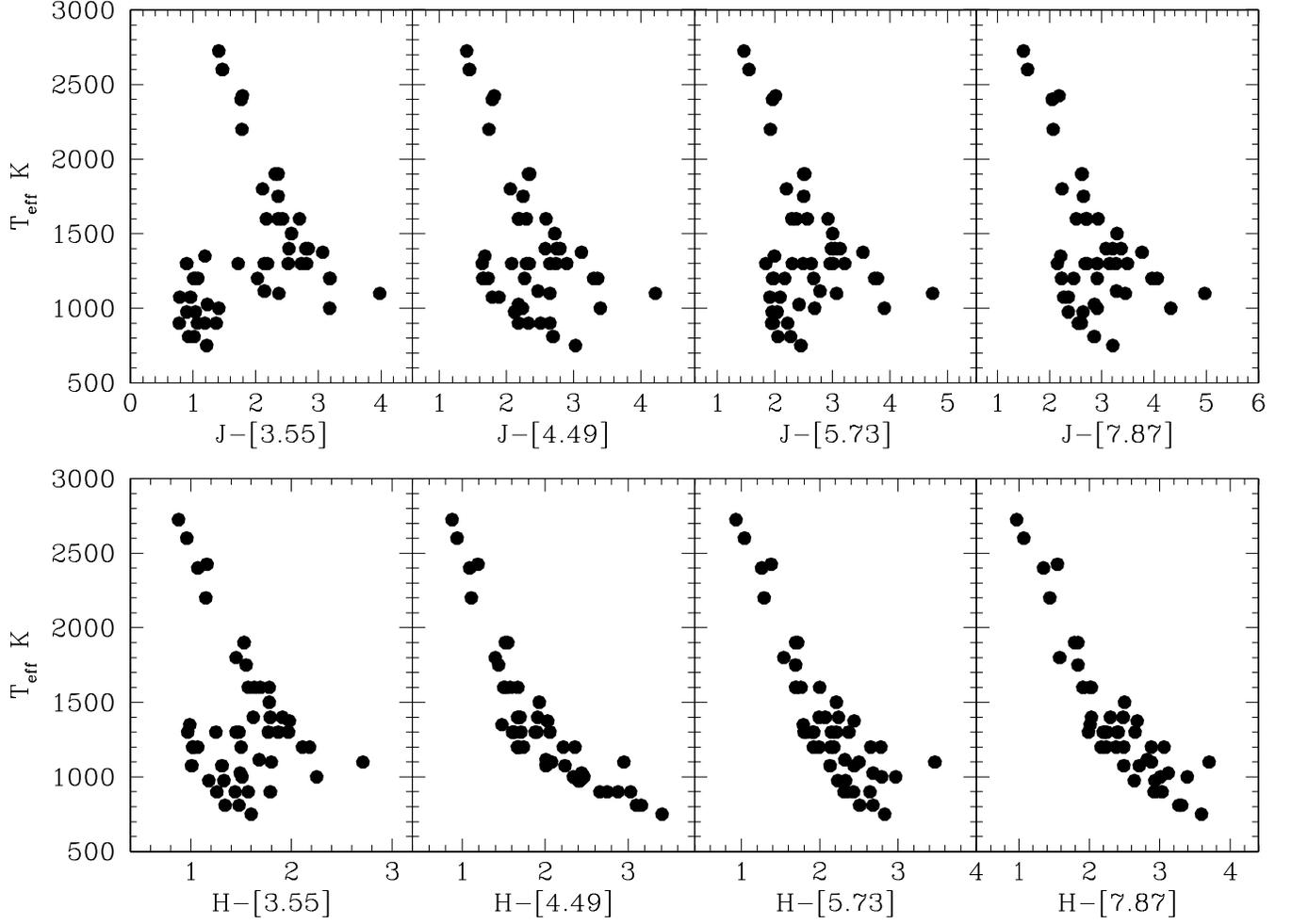}
\caption{Observed infrared colors of dwarfs constructed from the MKO $J$, MKO $H$, and IRAC bandpasses as a function of \teff.  The outlying point at 1100~K in each plot represents the very red L dwarf 2MASS~2244$+$20 (L7.5).  The values of \teff\ are from this paper or Golimowski et al.\ (2004a).  The IRAC photometry is taken from Patten et al.\ (2006) and Leggett et al.\ (2007).  The $J$ and $H$ photometry is taken from Leggett et al.\ (2002), Knapp et al.\ (2004), and Chiu et al.\ (2006). The uncertainties in \teff\ are 100 -- 300~K, and the uncertainties in the colors are 3 -- 10\%.
\label{fig14}}
\end{figure}

\clearpage

\begin{figure}
\includegraphics[angle=0,scale=.60]{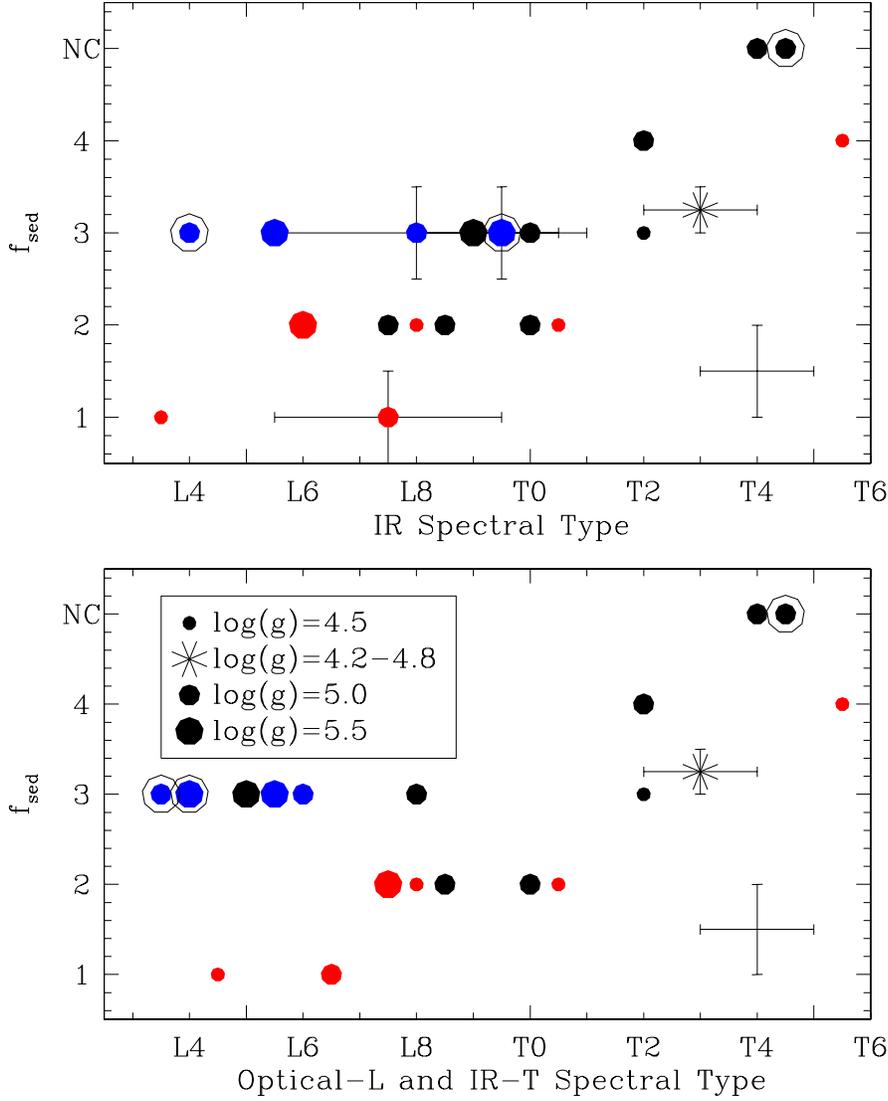}
\caption{Values of the grain sedimentation efficiency parameter, \fsed, as a function of infrared spectral type, top panel, and of optical type for the L dwarfs, lower panel.  Cloud-free or no-cloud models are indicated as `NC'. In order to separate datapoints both the optical and infrared type for DENIS~0255$-$47 have been changed by half a subclass to L8.5, and the optical type for  2MASS~1507$-$16 has also been changed by half a subclass, to L5.5.
The shapes and colors of the symbols are described in Figure~13.  The error bar shows the typical uncertainties in  \fsed\ and  spectral type. The smaller uncertainties in \fsed\ for HN Peg B are indicated, and the larger uncertainties in infrared spectral type for 2MASS 2244$+$20 (L7.5), SDSS 0805$+$48 (L9.5) and SDSS 1331$-$01  (L8) are indicated in the top panel.
\label{fig15}}
\end{figure}
\clearpage

\begin{figure}
\includegraphics[angle=-90,scale=.60]{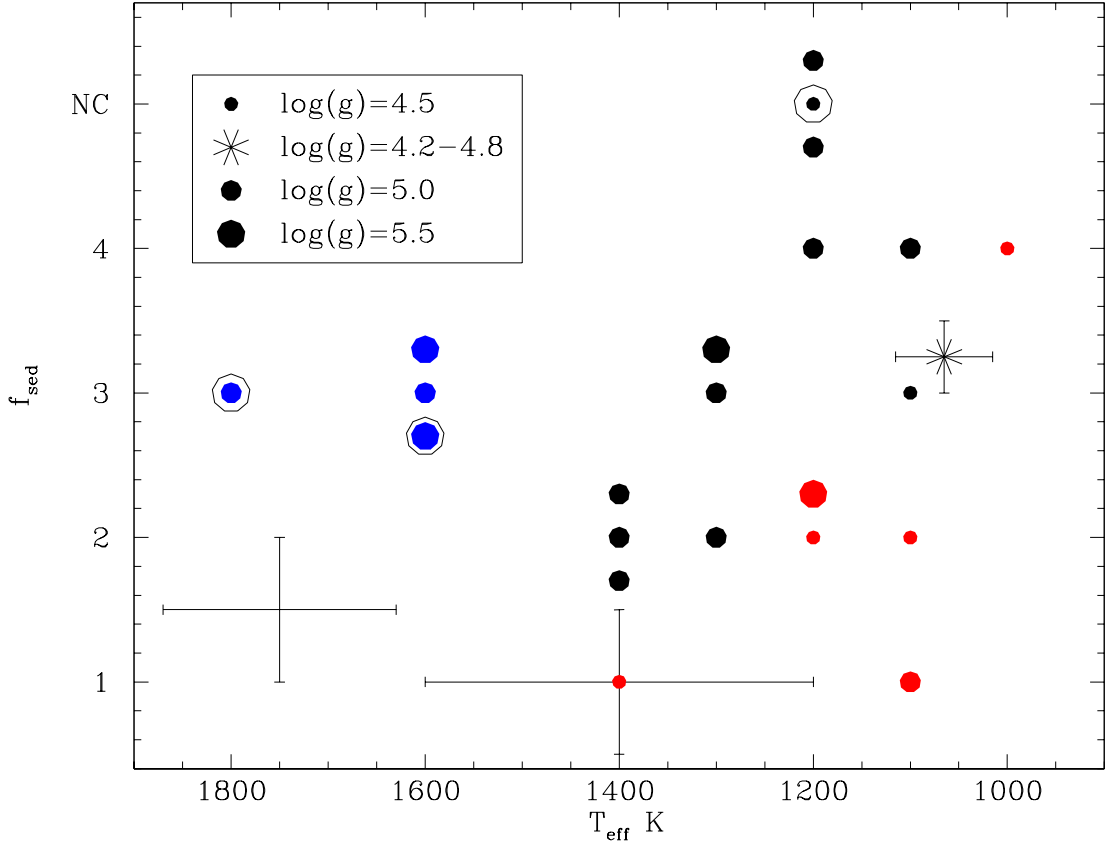}
\caption{Values of the grain sedimentation efficiency parameter, \fsed,  as a function of \teff. 
Cloud-free or no-cloud models are indicated as `NC'. In order to separate datapoints some values of \fsed\
have been changed by $\pm 0.3$. The shapes and colors of the symbols are described in Figure~13.  The error bar shows the typical uncertainties in  \fsed\ and  \teff.  The smaller uncertainties in both  \fsed\ and  \teff\ for HN Peg B are indicated, and the larger uncertainty in \teff\ for 2MASS~2224$-$01 (L3.5) is indicated. The parameters for the two components of  SDSS 1052$+$44AB are included in the plot.
\label{fig16}}
\end{figure}

\end{document}